\documentclass{jfm}
\usepackage{amsmath}
\RequirePackage{lfmath} 
\usepackage[sort&compress,longnamesfirst,colon,round]{natbib}
\usepackage{amsfonts}
\usepackage[pdftex,breaklinks]{hyperref}

\title[two-layer quasi-geostrophic turbulence]{The effect of asymmetric large-scale dissipation on energy and potential enstrophy injection in two-layer quasi-geostrophic turbulence}
\author{Eleftherios Gkioulekas}
\affiliation{University of Texas-Pan American, Department of Mathematics, Edinburg TX, 78539-2999, USA}

\begin{document}
\maketitle
\allowdisplaybreaks

\begin{abstract}
In the Nastrom-Gage spectrum of atmospheric turbulence we observe a $k^{-3}$ energy spectrum that transitions into a $k^{-5/3}$ spectrum, with increasing wavenumber $k$. The transition occurs near a transition wavenumber $k_t$, located near the Rossby deformation wavenumber $k_R$. The Tung-Orlando theory interprets this spectrum as a double downscale cascade of potential enstrophy and energy, from large scales to small scales, in which the downscale potential enstrophy cascade coexists with the downscale energy cascade over the same length-scale range. We show that, in a temperature forced two-layer quasi-geostrophic model, the rates with which potential enstrophy and energy are injected place the transition wavenumber $k_t$ near $k_R$. We also show that if the potential energy dominates the kinetic energy in the forcing range, then the Ekman term suppresses the upscale cascading  potential enstrophy more than it suppresses the upscale cascading  energy, a behavior contrary to what occurs in two-dimensional turbulence. As a result, the ratio $\gn/\gee$ of injected potential enstrophy over injected energy, in the downscale direction, decreases, thereby tending to decrease the  transition wavenumber  $k_t$  further.  Using a random Gaussian forcing model, we reach the same conclusion, under the modeling assumption that the asymmetric Ekman term predominantly suppresses the bottom layer forcing, thereby disregarding a possible entanglement between the Ekman term and the nonlinear interlayer interaction. Based on these results, we argue that the Tung-Orlando theory can account for the approximate coincidence between $k_t$ and $k_R$. We also identify certain open questions that require further investigation via numerical simulations.
\end{abstract}

\section{Introduction}

Quasi-geostrophic  models capture the dynamics of the atmosphere at  planetary scales greater than 100km, in order of magnitude. They are based on the assumptions of rapid rotation and small vertical thickness, both of which become  pronounced features of the dynamics of atmospheric motion  at increasingly larger length scales.  The simplest quasi-geostrophic model is the two-layer model, in which we have two layers of two-dimensional vorticity-streamfunction equations, coupled by a temperature equation, situated in a mid-layer between the vorticity layers. Obviously, the two-layer model can be generalized by adding more layers of vorticity-streamfunction equations interlaced with temperature equation mid-layers. In the limit of an infinite number of layers, we converge to the full quasi-geostrophic model. 

Until recently, it was assumed that quasi-geostrophic turbulence has the same dynamical behaviour as two-dimensional turbulence, where, according to the  theory of \citet{article:Kraichnan:1967:1}, \citet{article:Leith:1968}, and \citet{article:Batchelor:1969}, there is a downscale enstrophy cascade and an upscale inverse energy cascade. This assumption follows from \citet{article:Charney:1971}  who claimed that there is an ``isomorphism'' between quasi-geostrophic turbulence and two-dimensional turbulence. More recently, key differences between these two models were noted by \cite{article:Welch:2001}, \citet{article:Orlando:2003:1}, and \citet{article:Tung:2007}. The most remarkable difference was highlighted in the numerical simulation of the two-layer quasi-geostrophic model by \citet{article:Orlando:2003},  which produced an energy spectrum that scales as  $k^{-3}$  initially, and with increasing wavenumber $k$,  transitions to  $k^{-5/3}$  scaling. This is  consistent with the observed energy  spectrum of the atmosphere, as was first measured by \citet{article:Gage:1984} and \citet{article:Nastrom:1986}, but it is inconsistent with our conventional understanding of the dynamical behavior of two-dimensional turbulence, as described by the theory of \citet{article:Kraichnan:1967:1}, \citet{article:Leith:1968}, and \citet{article:Batchelor:1969}. \citet{article:Orlando:2003} showed that their simulation produced a  downscale enstrophy cascade that co-existed with a downscale energy cascade, with  both energy and enstrophy  injected by baroclinic instability at small wavenumbers, and dissipated at large wavenumbers. If $\gn$ is the enstrophy flux and $\gee$  is the energy flux associated with these coexisting cascades, then, by dimensional analysis, \citet{article:Orlando:2003} argue that the transition from  $k^{-3}$ scaling to $k^{-5/3}$ scaling should occur at a transition wavenumber $k_t \sim (\gn/\gee)^{1/2}$, and it does.

This result bolstered the Tung-Orlando theory that interpreted the Nastrom-Gage spectrum as a coexisting downscale potential enstrophy cascade and downscale energy cascade, both spanning a comparable range of length scales. It should be noted that it was not the intention of \citet{article:Orlando:2003}  to claim that the entire Nastrom-Gage spectrum can be explained via the two-layer quasi-geostrophic model. The  point of the simulation was to demonstrate that it is possible to have coexisting downscale potential enstrophy and energy cascades, even in models as close to two-dimensional turbulence as the two-layer quasi-geostrophic model. This possibility is bound to become even more favorable under models that are further away from the two-dimensional approximation, such as the multi-layer quasi-geostrophic model or the three-dimensional stratified turbulence model. Gravity waves can also play a helpful role in facilitating coexisting cascades, as discussed further in section 5. 

In a subsequent paper, \citet{article:Smith:2004} criticized \citet{article:Orlando:2003}  on the following grounds: First,  using his ``HVK scale'' estimates, he claimed that the Tung-Orlando numerical simulation is not well-resolved and that therefore the $k^{-5/3}$  part of the Tung-Orlando energy spectrum is a bottleneck instead of being indicative of a real cascade. In connection with this claim,  \citet{article:Smith:2004} criticized the use of a resolution-dependent hyperdiffusion coefficient by  \citet{article:Orlando:2003}. Second, that in two-dimensional turbulence it is not possible for the downscale energy flux to be large enough to create a gap between the transition wavenumber $k_t$ and the dissipation wavenumber $k_d$. In  response, \citet{article:Tung:2004} noted that: (a) Since, the diagnostics in the Tung-Orlando simulation indicate a downscale enstrophy flux $\gn$  and a downscale energy flux $\gee$   that have magnitudes consistent with the location of the transition wavenumber $k_t$  in the simulation's energy spectrum, it is very unlikely that the transition is caused by an energy bottleneck, as argued by \citet{article:Smith:2004}. (b) The use of a resolution-dependent hyperviscosity coefficient is intended to model the anomalous energy dissipation sink at small scales, originating from three-dimensional dynamics, by controlling the downscale energy dissipation rate. (c)  Smith's ``HVK scale'' argument, which was used to argue that the Tung-Orlando simulation is not well-resolved, has various flaws, discussed in detail by \citet{article:Tung:2004}, that render it inconclusive. Nevertheless, Smith's claim, that $k_t$ will coincide with the dissipation scale $k_d$, can be shown to hold, for the case of two-dimensional turbulence, via a corrected proof given by \citet{article:Tung:2005:1}. However, as was shown by \cite{article:Tung:2007}, this result is not necessarily generalizable to quasi-geostrophic models. Thus, \citet{article:Smith:2004}  did not establish the claim that ``an inertial range transition is not possible in quasi-geostrophic models'',  and the theoretical problem remains open. (d) Since the diagnostics of the Tung-Orlando simulation indicate that the downscale energy dissipation rate balances the rate with which energy is sent downscale from the forcing range, the simulation is sufficiently well-resolved to prevent a bottleneck-type energy pile up at small scales, over a time-scale longer than the runtime of the simulation.

Be that as it may, the underlying theoretical question, implied by \citet{article:Smith:2004}, remained open: How can the downscale energy flux  $\gee$   be large enough to yield a gap between $k_t$  and  $k_d$, when that is not possible in two-dimensional turbulence? An even deeper question also demanded further understanding: how is it possible for two downscale cascades to coexist? These lingering questions generated skepticism towards the Tung-Orlando theory, which is why we were prompted to investigate them at greater depth. In \citet{article:Tung:2005,article:Tung:2005:1}, we augmented the Tung-Orlando theory by noting that even in two-dimensional turbulence  there is  a small amount of energy cascading from small to large wavenumbers, as long as the viscosity coefficient of the small-scale dissipation term is non-zero. We have proposed that this small ``energy leak'' should be viewed as a downscale energy cascade that coexists with the dominant downscale enstrophy cascade. To support this theory, in \citet{article:Tung:2005,article:Tung:2005:1} we noted that  the triad interactions responsible for the enstrophy cascade are independent from those responsible for the downscale energy cascade. This is, in fact, an immediate but unstated consequence of the original argument by \citet{article:Kraichnan:1967:1}, as noted in section 3.2 of \citet{article:Tung:2005}.  We have also theorized that the two cascades can be viewed as two independent homogeneous solutions of the governing statistical theory that can be linearly superposed on each other. \citet{article:Davidson:2008}  confirmed the validity of the linear superposition principle for the 3rd-order structure functions, thereby adding further detail to a corresponding proof sketch given in section 3.1 of \citet{article:Tung:2005}. Some of the details of my statistical theory of two-dimensional turbulence was given in \citet{article:Gkioulekas:2008:1} and \citet{article:Gkioulekas:p14}, and further development of this theory is currently in progress.

To elaborate further, our claim is that the energy spectrum of the downscale cascade is given by the linear combination of a dominant $k^{-3}$  term, arising from the dominant downscale enstrophy cascade, and a subdominant $k^{-5/3}$  term, arising from the hidden downscale energy cascade, which allows, in principle, a transition from the $-3$ slope to the $-5/3$ slope.  In linearly dissipated two-dimensional turbulence,  this transition is not expected to be realized, because an upper bound  on the energy flux  forces the transition scale $k_t$ to be greater than the dissipation scale $k_d$ of the enstrophy cascade. If $\Pi_E (k)$ is the energy flux from the $(0,k)$ wavenumber interval to the $(k,+\infty)$  interval and  $\Pi_G (k)$   is the enstrophy flux from  $(0,k)$ to  $(k,+\infty)$, then this flux inequality reads $k^2 \Pi_E (k)-\Pi_G (k)<0$, for all wavenumbers $k$  not in the forcing range. The origin of this inequality is  the relationship $D_E (k) = k^{-2} D_G (k)$ between  the  energy dissipation spectrum $D_E (k)$ and  the enstrophy dissipation spectrum $D_G (k)$.  Thus, with increasing wavenumber $k$, the proportion of the energy dissipation rate relative to the enstrophy dissipation rate vanishes rapidly with $k^{-2}$, and this rapid decrease is the reason why the downscale energy cascade coexisting with the downscale enstrophy cascade cannot be seen in the energy spectrum. However, the subtle point  that deserves to be stressed here is that \emph{the proof of the inequality  involves both the linearity of the dissipation terms and the twin conservation laws (of enstrophy and energy), and is not inherent solely to the twin conservation laws.} With a nonlinear dissipation term, that could result from neglected three-dimensional effects, the  flux  inequality could be violated within the inertial range. Since the transition wavenumber $k_t$ is  expected to be approximately equal to the wavenumber where the flux inequality becomes an equation, an inertial range violation of the flux inequality would give $k_t \ll k_d$. 

As was shown by \citet{article:Tung:2007}, in the quasi-geostrophic two-layer model, the relationship between  $D_E (k)$  and   $D_G (k)$ becomes quite complicated, so it may be possible to violate the flux inequality, thus resulting  in a significant separation between $k_t$ and $k_d$.  If that occurs, we can expect  $k^{-5/3}$ scaling in the gap created between $k_t$ and $k_d$. In \citet{article:Tung:2007}, we have identified asymmetric dissipation as the only mechanism that can break the flux inequality. By asymmetric dissipation we mean that the dissipation operators acting on the top and bottom velocity equations are different: indeed, in the two-layer quasi-geostrophic model there is  an Ekman  dissipation term  acting at large scales at the bottom layer but not at the top layer. Since the small-scale hyperdiffusion is not physically inherent in the quasi-geostrophic dynamics, there is no physical reason to prefer symmetric over asymmetric hyperdiffusion either. Unfortunately, there are still many open questions concerning the theory of the flux inequality. Consequently, the numerical results of \citet{article:Orlando:2003} notwithstanding, there is still some uncertainty on whether the two-layer model can have a robust energy dissipation sink that can break the flux inequality in the inertial range. 

It should be noted that when the same dissipation operator is  used on both layers, it can be proved that the flux inequality is satisfied for all wavenumbers not in the forcing range. For that case, the prediction of \citet{article:Charney:1971}, that quasi-geostrophic turbulence will be isomorphic to two-dimensional turbulence, is expected to hold. This was confirmed  in a numerical simulation by \citet{article:Lindborg:2010}, where the dissipation operator and the forcing term are both independent of the vertical coordinate. 

Recently, \citet{article:Smith:2009} proposed a more sophisticated two-layer two-mode quasi-geostrophic model that has succeeded in reproducing the Nastrom-Gage spectrum. More importantly, using their quasi-geostrophic model, \citet{article:Smith:2009} confirmed that it is possible for a downscale potential enstrophy cascade to coexist with a downscale energy cascade, thereby vindicating the fundamental premise of the Tung-Orlando theory.  A further advantage of the Tulloch-Smith QG model is that it seems to have resolved the small-scale energy dissipation sink problem implied by the HVK argument of \citet{article:Smith:2004}. Since the Tulloch-Smith QG model is still a severely restricted approximation of the full quasi-geostrophic model, it is reasonable to expect that the small-scale energy dissipation sink problem will remain resolved under  the full quasi-geostrophic model. This raises many interesting questions that are, unfortunately, beyond the scope of this paper, but will be investigated in future work.

The goal of the present paper is to add one more piece to the overall puzzle by looking at the forcing range instead of the dissipation range.  We will thus consider the effect of symmetric versus asymmetric forcing on the dynamics of the two-layer quasi-geostrophic model. We will first show that when the model is forced exclusively through the temperature equation, this results in antisymmetric forcing on the potential vorticity equations for both layers. Consequently,  the energy forcing spectrum $F_E(k)$ and the potential enstrophy forcing spectrum $F_G(k)$ are related as  $F_G(k) = (k^2+k_R^2) F_E(k)$, with $k_R$ the Rossby wavenumber.   For forcing-range wavenumbers $k\sim k_f \ll k_R$, we have approximately $F_G(k) \sim k_R^2 F_E(k)$. It follows that if we neglect Ekman dissipation, then the ratio of the enstrophy flux $\gn$ over the energy flux $\gee$ injection to the downscale cascades  will satisfy $(\gn/\gee) \sim k_R^2$, and therefore the transition wavenumber $k_t$ has to be in the vicinity of the Rossby wavenumber $k_R$. As it turns out, this is indeed the approximate location of the transition wavenumber  $k_t$ in the actual Nastrom-Gage spectrum as well as in the \citet{article:Orlando:2003} simulation.

We will show that asymmetric Ekman dissipation tends to decrease the ratio $k_t \sim (\gn/\gee)^{1/2}$ as long as the potential energy spectrum dominates the kinetic energy spectrum  in the forcing range. This peculiar behaviour results from the asymmetry of the effective forcing between the two layers, caused by the introduction of the Ekman term into the bottom layer. This claim is further supported by our consideration of the random Gaussian forcing model, in which the bottom-layer forcing is directly suppressed by a controlled scalar factor. Unfortunately, there are a number of open questions and outstanding issues concerning the distribution of energy between potential energy and kinetic energy. Furthermore, the modeling assumption that the Ekman term suppresses forcing only at the lower-layer is equivalent to ignoring the unknown effect of the entanglement of the Ekman term with the interlayer interaction, and that  is the underlying problem.

It should be noted that, in the context of the two-layer model, unless the dissipation terms at small scales can dissipate the energy and potential enstrophy at the same rate with which they are injected to the downscale range,  the downscale energy and potential enstrophy cascades will simply fail to develop. It is not yet obvious, in terms of theory, whether the two-layer quasi-geostrophic model can dissipate this much energy, a problem previously discussed by \citet{article:Tung:2007}. On the other hand, in the real atmosphere, we note that at larger wavenumbers, the dynamics  transitions from quasi-geostrophic  to stratified three-dimensional turbulence. According to \citet{article:Lindborg:2007}, the transition to stratified turbulence occurs at a scale of about $100$km. Since stratified turbulence, like three-dimensional Navier-Stokes turbulence, does have an anomalous energy dissipation sink, it follows that any amount of energy injected at large scales can and will be dissipated. Furthermore, since potential enstrophy continues to be conserved under stratified dynamics, the two cascades can continue to coexist for scales less than $100$km. On the other hand,  the two-layer model is indeed realistic at the small synoptic-scale wavenumbers, where the forcing takes place, so using it to explain the rates of energy and potential enstrophy injection at the forcing range is a fair argument. 

As we have explained above, in the Tung-Orlando theory, the location  of the transition wavenumber $k_t$ is directly determined by the relative magnitude of the downscale potential enstrophy flux $\gn$ over the downscale energy flux  $\gee$. A different mechanism underlies an SQG model that was recently proposed by \citet{article:Smith:2006}. In their model, there is only one cascade, whose scaling exponent changes with wavenumber $k$, because the self-similar scaling of the model's nonlinear term changes with increasing $k$.  As a result, in the Tulloch-Smith SQG model the transition wavenumber  $k_t$  is strictly constrained to coincide with  $k_R$, because the transition in the scaling of the nonlinear term of the model occurs at $k_R$.  This dynamics of the Tulloch-Smith SQG  model is analogous to that of the LANS $\ga$-model of three-dimensional Navier-Stokes turbulence  \citep{article:Wingate:2005}, in which, once again there is a single downscale energy cascade with $k^{-5/3}$ scaling in the energy spectrum, with a transition to  a steeper $k^{-3}$ slope at higher wavenumbers $k$, because of the introduced  distortion of the Navier-Stokes nonlinearity. In both cases, we are essentially dealing with a single homogeneous solution, associated with a single flux coefficient, which, to first approximation, we can consider bifractal.  

Under the quasi-geostrophic models used by \citet{article:Orlando:2003} and \citet{article:Smith:2009},  on the other hand, we are dealing with two independent homogeneous solutions each of which, to first approximation, can be considered monofractal.  Because the two solutions are independent of each other, as far as the nonlinearity is concerned, it is possible for the transition wavenumber  $k_t$, in principle, to have any arbitrary value, since its location is determined solely by the relative magnitude of the two homogeneous solutions against each other. From the viewpoint of this paper,  the constraint $k_t \sim k_R$  is a weak indirect constraint that originates from the combined effect of anti-symmetric forcing and the large-scale Ekman dissipation term on the energy and potential enstrophy injection rates. This paper argues that the two-layer quasi-geostrophic model is consistent with placing $k_t$ near $k_R$. 

The paper is organized as follows. In section 2 we review the two-layer quasi-geostrophic model and introduce various preliminaries. In section 3 we derive the potential enstrophy and energy forcing spectra for the case of a generalized multi-layer quasi-geostrophic model, and discuss the random Gaussian forcing model. These results are applied to the two-layer quasi-geostrophic model itself in section 4. Conclusions and discussion are given in section 5 and technical matters are discussed in the appendices.

\section{The two-layer model equations}

The two-layer model is defined by two vorticity-streamfunction equations and a temperature equation which read
\begin{align}
\pderiv{\gz_1}{t} &+ J(\gy_1, \gz_1+f) = -\frac{2f}{h} \gw + d_1, \label{eq:RelativeVortOne} \\
\pderiv{\gz_2}{t} &+ J(\gy_2, \gz_2+f) = +\frac{2f}{h} \gw + d_2 + e_2, \label{eq:RelativeVortTwo}  \\
\pderiv{T}{t} &+ J\left( \frac{\gy_1+\gy_2}{2}, T\right)  = -\frac{N^2}{f}\gw + Q_0. \label{eq:Temperature} 
\end{align}
We see that the temperature $T$ is advected by the average streamfunction $(\gy_1+\gy_2)/2$. Here  $\gy_1$  and  $\gy_2$  are the streamfunctions of the top and bottom layers; $\gw$ is the vertical velocity; $\gz_1 = \del^2 \gy_1$ and $\gz_2 = \del^2 \gy_2$ are the relative vorticities, and $d_1$, $d_2$, and $e_2$ are the dissipation terms given by 
\begin{align}
d_1 &= (-1)^{\gk+1}\nu \del^{2\gk} \gz_1 =  (-1)^{\gk+1}\nu \del^{2\gk+2} \gy_1 ,\\
d_2 &= (-1)^{\gk+1}\nu \del^{2\gk} \gz_2 =  (-1)^{\gk+1}\nu \del^{2\gk+2} \gy_2, \\
e_2 &= -\nu_E \gz_2 = -\nu_E \del^2 \gy_2.
\end{align}
The terms $d_1$ and $d_2$ represent  momentum dissipation of relative vorticity and $e_2$ represents Ekman damping from the lower boundary layer. Furthermore, $h$ is the height between the top and bottom rigid horizontal boundaries (the two vorticity layers and the temperature midlayer divide the space between the horizontal boundaries into four equal intervals, with the temperature midlayer situated between the two vorticity layers), $f$ is the Coriolis term, $N$ is the  \BVdude  frequency, and $Q$ the thermal forcing term. The temperature $T$ is related with the streamfunctions  $\gy_1$  and  $\gy_2$ via the geostrophic condition  $T = (2/h)(\gy_1-\gy_2)$. Finally, $J(a, b)$ is defined as the Jacobian between the fields $a$ and $b$ and it reads:
\begin{equation}
J (a,b) = \pderiv{a}{x} \pderiv{b}{y} - \pderiv{b}{x} \pderiv{a}{y}.
\end{equation}

Solving for the vertical velocity $\gw$ in the temperature equation, after substituting the geostrophic condition, leads to the definition of the potential vorticities $q_1$ and $q_2$  given by
\begin{align}
q_1 &=  \del^2 \gy_1 + f + \frac{k_R^2}{2}(\gy_2-\gy_1), \label{eq:DefPotVortOne} \\
q_2 &=  \del^2 \gy_2 + f - \frac{k_R^2}{2}(\gy_2-\gy_1), \label{eq:DefPotVortTwo} 
\end{align}
and their corresponding governing equations which read:
\begin{align}
\pderiv{q_1}{t} &+ J(\gy_1, q_1) = f_1 + d_1, \label{eq:PotVortOne}  \\
\pderiv{q_2}{t} &+ J(\gy_2, q_2) = f_2 + d_2+ e_2. \label{eq:PotVortTwo}
\end{align}
Here, $k_R$ is the Rossby deformation wavenumber defined as $k_R \equiv 2\sqrt{2} f/(hN)$,  $f_1 = -(1/4)k_R^2 h Q_0$, and $f_2 = (1/4)k_R^2 h Q_0$. The derivation is shown in Appendix~\ref{app:VorticityStreamfunctionEquations}. Although the argument is well-known folklore, we want to note mainly that: (a) the dissipation terms have the same form in the relative vorticity equations as they do in the potential vorticity equations; (b) the thermal forcing term $Q$ appears on both top and bottom potential vorticity equations with opposite signs. Consequently, both layers are forced anti-symmetrically by the same forcing term, except with opposite signs.

It is also well-known that the two-layer model, in the absence of forcing and dissipation, conserves the total energy E given by 
\begin{equation}
E(t) = - \int_{\bbR^2} [\psi_1 (\bfx,t) q_1 (\bfx,t)  + \psi_2 (\bfx,t)  q_2 (\bfx,t) ] \;\df{\bfx},
\end{equation}
and the potential enstrophies $G_1$ and $G_2$ for each layer given by:
\begin{align}
G_1 (t) &= \int_{\bbR^2} q_1^2  (\bfx,t)  \;\df{\bfx},\\
G_2 (t) &= \int_{\bbR^2} q_2^2  (\bfx,t)  \;\df{\bfx}.
\end{align}

To properly define all the relevant spectra associated with these conserved quantities, consider first the Fourier expansions of the streamfunctions fields $\psi_\ga (\bfx,t)$ and potential vorticity fields $q_\ga (\bfx,t)$  ($\ga = 1,2$):
\begin{align}
\psi_\ga (\bfx,t) &= \int_{\bbR^2} \hat\psi_\ga (\bfk,t) \exp (i\bfk\cdot\bfx) \; \df{\bfk}, \\
q_\ga (\bfx,t) &= \int_{\bbR^2} \hat q _\ga (\bfk,t) \exp (i\bfk\cdot\bfx) \; \df{\bfk}.
\end{align}
In Fourier space, the potential vorticity  fields $q_\ga$ and streamfunction  fields $\psi_\ga$ are related by 
\begin{equation}
\hat q _\ga (\bfk,t) = \sum_{\gb} L_{\ga\gb} (\nrm{\bfk}) \hat\psi_\ga (\bfk,t). \label{eq:QpsiRel}
\end{equation}
Here, the sum runs over all layers, in this case $\gb=1,2$, and  $L_{\ga\gb} (k)$ is a wavenumber matrix defined as
\begin{equation}
L_{\ga\gb} (k) = \mattwo{-k^2-k_R^2/2}{+k_R^2/2}{+k_R^2/2}{-k^2-k_R^2/2}.
\label{eq:TheMatrixL}
\end{equation}
In real space, the same relation between the potential vorticity $q_\ga$ and the streamfunction $\psi_\ga$ can be written in terms of a corresponding differential operator $\ccL_{\ga\gb}$, as follows:
\begin{equation}
q_\ga (\bfx,t) = \sum_{\gb} \ccL_{\ga\gb} \psi_\ga (\bfx,t).
\end{equation}
It is easy to see that the matrix $L_{\ga\gb} (k)$  is non-singular, for $k>0$, and can therefore be inverted. The inverse matrix $L_{\ga\gb}^{-1} (k)$ defines a corresponding inverse integrodifferential operator $\ccL_{\ga\gb}^{-1}$. Note that in Eq.~\eqref{eq:QpsiRel} we have neglected the $\gb$ contribution to the Coriolis term $f$, since, for the case of our planet, the impact of the $\gb$-effect on the Nastrom-Gage energy spectrum is negligible. We have also neglected the latitude dependence of $f$, on the premise that we are interested in the ensemble average of the energy spectrum restricted on a thin strip of the Earth's surface that is oriented parallel to the equator. These approximations cause the Coriolis term $f$ to  drop out of the nonlinear Jacobian terms altogether.

Let us now introduce the following notation. Consider any arbitrary abstract scalar fields $a(\bfx)$ and $b(\bfx)$, which can be snapshots in time of either the streamfunction fields $\psi_\ga (\bfx,t)$  or the potential vorticity fields  $q_\ga (\bfx,t)$ for a given level $\ga$. Let $a^{<k}(\bfx)$ and $b^{<k}(\bfx)$ be the fields obtained from $a(\bfx)$ and $b(\bfx)$ by setting to zero, in Fourier space, the components corresponding to wavenumbers whose norm is greater than $k$. Formally, $a^{<k}(\bfx)$ is defined as
\begin{align}
a^{<k}(\bfx) &= \int_{\bbR^2} \df{\bfx_0} \int_{\bbR^2} \df{\bfk_0}\;  \frac{H(k-\nrm{\bfk_0})}{4\pi^2} \exp (i\bfk_0\cdot (\bfx-\bfx_0)) a(\bfx_0) \\
&\equiv \int_{\bbR^2} \df{\bfx_0} \; P(k|\bfx-\bfx_0) a(\bfx_0),
\end{align}
with $H(x)$ the Heaviside function, defined as the integral of a delta function:
\begin{align}
H(x) &= \int_0^x \gd (\tau) \; d\tau = \casethree{1}{\text{if } x\in (0,+\infty)}{1/2}{\text{if } x=0}{0}{\text{if } x\in (-\infty,0)}.
\end{align}
Here $P(k|\bfx-\bfx_0)$ is the corresponding low-pass filter kernel. Obviously, $b^{<k}(\bfx)$ is defined similarly. We now use the two filtered fields $a^{<k}(\bfx)$  and $b^{<k}(\bfx)$ to define the bracket $\innerf{a}{b}{k}$  as: 
\begin{align}
\innerf{a}{b}{k} &= \dD{k} \int_{\bbR^2} \df{\bfx}\;\avg{a^{<k}(\bfx) b^{<k}(\bfx)} \label{eq:BracketDef} \\
&= \frac{1}{2} \int_{A\in \SO{2}} \df{\Omega (A)}\; \avg{[\hat a^{\ast}(kA\bfe) \hat b(kA\bfe) + \hat a(kA\bfe) \hat b^{\ast}(kA\bfe)]}. \label{eq:BracketFourier} 
\end{align}
Here, $\hat a (\bfk)$ and $\hat b (\bfk)$ are the Fourier transforms of $a(\bfx)$ and $b(\bfx)$, $\SO{2}$ is the set of all non-reflecting rotation matrices in two dimensions, $d\Omega (A)$ is the measure of a spherical integral,  $\bfe$ is a two-dimensional unit vector, and $\avg{\cdot}$  represents taking an ensemble average. The star superscript represents taking the complex conjugate. Also note that Eq.~\eqref{eq:BracketDef} is the definition of the bracket, and Eq.~\eqref{eq:BracketFourier}  is a consequence of the definition. 

It is easy to see that the bracket is both symmetric and bilinear, in that it satisfies:
\begin{align}
&\innerf{a}{b}{k} = \innerf{b}{a}{k},\\
&\innerf{a}{b+c}{k} = \innerf{a}{b}{k} + \innerf{a}{c}{k}, \\
&\innerf{a+b}{c}{k} = \innerf{a}{c}{k} + \innerf{b}{c}{k}.
\end{align}
Moreover, every $(\ab)$-component of the operator $\ccL_{\ga\gb}$ is self-adjoint with respect to the bracket, which gives
\begin{equation}
\innerf{\ccL_{\ab} a}{b}{k}  = \innerf{a}{\ccL_{\ab} b}{k} = L_{\ab} (k)\innerf{a}{b}{k},
\end{equation}
and the same property is also satisfied by every component of the inverse operator $\ccL_{\ga\gb}^{-1}$:
\begin{equation}
\innerf{\ccL_{\ab}^{-1} a}{b}{k}  = \innerf{a}{\ccL_{\ab}^{-1} b}{k} = L_{\ab}^{-1} (k)\innerf{a}{b}{k}.
\end{equation}

Using the bracket, we now define the energy spectrum $E(k)= - \innerf{\psi_1}{q_1}{k} - \innerf{\psi_2}{q_2}{k}$ and the potential enstrophy spectra $G_1(k)= \innerf{q_1}{q_1}{k}$ and $G_2(k)= \innerf{q_2}{q_2}{k}$ for each layer. We also define $G (k)=G_1(k)+G_2(k)$  as the total potential enstrophy spectrum. This method of defining spectra was previously used by \citet{book:Frisch:1995}, and it is equivalent to the standard definition of spectra in terms of narrow spherical shells in Fourier space (see Eq.~\eqref{eq:BracketFourier}). It is also superior in that one can generalize the definition of spectra to non-homogeneous flows by removing the spatial integral in Eq.~\eqref{eq:BracketDef}, thereby obtaining location-dependent spectra.

It is useful to be able to rewrite the above spectra in terms of a streamfunction spectrum $C_{\ga\gb}(k)=\innerf{\psi_\ga}{\psi_\gb}{k}$. From the bilinear property of the bracket, it follows that the energy spectrum $E(k)$ reads:
\begin{align}
E(k) &= -\sum_{\ga}  \innerf{\psi_\ga}{q_\ga}{k} = -\sum_{\ga} \innerf{\psi_\ga}{\sum_{\gb} \ccL_{\ab} \psi_\gb}{k} = -\sum_{\ab} L_{\ab}(k) \innerf{\psi_\ga}{\psi_\gb}{k} \\
&= -\sum_{\ab} L_{\ab}(k) C_{\ab}(k). \label{eq:Ek}
\end{align}
Likewise, the potential enstrophy spectrum $G(k)$ reads:
\begin{align}
G (k) &= \sum_{\ga} \innerf{q_\ga}{q_\ga}{k} = \sum_{\ga} \innerf{\sum_{\gb} \ccL_{\ab} \psi_\gb}{\sum_{\gc} \ccL_{\ac} \psi_\gc}{k} \\ 
&= \sum_{\ab} L_{\ab} (k)\innerf{\psi_\gb}{\sum_{\gc} \ccL_{\ac} \psi_\gc}{k} =  \sum_{\abc} L_{\ab} (k) L_{\ac}(k) \innerf{\psi_\gb}{\psi_\gc}{k}\\
&= \sum_{\abc} L_{\ab} (k) L_{\ac}(k) C_{\bc}(k). \label{eq:Gk}
\end{align}
Stated in this way, these expressions easily generalize to multiple-layer quasi-geostrophic models simply by using a different matrix $L_{\ab} (k)$ with more rows and columns. 

\section{Forcing Spectrum in general}

Now let us turn our attention to the forcing spectra of the two-layer model. We begin by writing the potential vorticity equations in the following more abstract form:
\begin{equation}
\pderiv{q_\ga}{t}+J(\psi_\ga, q_\ga) = \sum_\gb \cD_{\ga\gb} \psi_\gb + f_\ga.
\end{equation}
Here, $\cD_{\ga\gb}$  is a matrix operator that accounts for all the dissipation terms and $f_\ga$ is the forcing term acting on the $\ga$-layer. Using this abstract formulation will shorten the calculations below considerably. For the case of thermal forcing, the forcing terms take the form $f_1 = \gf$ and $f_2 = -\gf$. 

Multiplying both sides with the inverse operator $\ccL_{\ga\gb}^{-1}$ gives  the following governing equation for the streamfunctions:
\begin{equation}
\pderiv{\psi_\ga}{t}+ \sum_\gb \ccL_{\ab}^{-1} J(\psi_\gb, q_\gb) = \sum_{\bc} \ccL_{\ab}^{-1} \cD_{\bc} \psi_\gc + \sum_\gb \ccL_{\ab}^{-1} f_\gb. \label{eq:GovEqStreamFunc}
\end{equation}
Now, let us define a streamfunction-forcing spectrum $\phi_{\ab}(k) = \innerf{f_\ga}{\psi_\gb}{k}$ and recall our definition of the streamfunction spectrum $C_{\ga\gb}(k)=\innerf{\psi_\ga}{\psi_\gb}{k}$. Differentiating $C_{\ab}(k)$ with respect to time gives
\begin{equation}
\pderiv{C_{\ab}(k)}{t} = \innerf{\pderiv{\psi_\ga}{t}}{\psi_\gb}{k} + \innerf{\psi_\ga}{\pderiv{\psi_\gb}{t}}{k}, \label{eq:TimeDerCab}
\end{equation}
and we may write a governing equation for $C_{\ab}(k)$ in the form
\begin{equation}
\pderiv{C_{\ab}(k)}{t} + T_{\ab}(k) = D_{\ab}(k) + F_{\ab}(k). \label{eq:GovEqCab}
\end{equation}
Here, $T_{\ab}(k)$  is the contribution from the nonlinear Jacobian term, $D_{\ab}(k)$  is the contribution from the dissipation term, and $F_{\ab}(k)$  is the contribution from the forcing term. Our interest here is in the forcing contribution $F_{\ab}(k)$. It is easy to see that $F_{\ab}(k)$ can be written in terms of the  streamfunction-forcing spectrum $\phi_{\ab}(k)$ as follows:
\begin{align}
F_{\ab}(k) &= \innerf{\sum_\gc \ccL_{\ac}^{-1} f_\gc}{\psi_\gb}{k} + \innerf{\psi_\ga}{\sum_\gc \ccL_{\bc}^{-1} f_\gc}{k} \\
 &= \sum_\gc L_{\ac}^{-1}(k) \innerf{f_\gc}{\psi_\gb}{k} + \sum_\gc L_{\bc}^{-1}(k) \innerf{\psi_\ga}{f_\gc}{k} \\
 &= \sum_\gc [ L_{\ac}^{-1}(k) \phi_{\cb} (k) + L_{\bc}^{-1}(k) \phi_{\ca} (k)].
\end{align}
Here, we have replaced the time derivatives in Eq.~\eqref{eq:TimeDerCab} with the forcing term from Eq.~\eqref{eq:GovEqStreamFunc}. We now note that governing equations for the energy spectrum $E(k)$ and  the potential enstrophy spectrum $G(k)$ can be obtained by applying the operators indicated by equations \eqref{eq:Ek} and \eqref{eq:Gk} to the governing equation for the streamfunction spectrum $C_{\ab}(k)$. These equations are analogous to Eq.~\eqref{eq:GovEqCab}  and they take the form:
\begin{align}
&\pderiv{E(k)}{t} + T_E (k) = D_E (k) + F_E (k),\\
&\pderiv{G(k)}{t} + T_G (k) = D_G (k) + F_G (k).
\end{align}
Here, the terms above have analogous definitions.  The next step is to write the forcing spectrum $F_E (k)$ for the energy and  $F_G (k)$  for the potential enstrophy in terms of   $F_{\ab}(k)$. Using the operator indicated by Eq. \eqref{eq:Ek}, the energy forcing spectrum $F_E (k)$ reads:
\begin{align}
F_E (k) &= -\sum_{\ab} L_{\ab}(k) F_{\ab}(k) = -\sum_{\abc} L_{\ab}(k) [ L_{\ac}^{-1}(k) \phi_{\cb} (k) + L_{\bc}^{-1}(k) \phi_{\ca} (k)] \\
&= -\sum_{\bc} \left[  \left(  \sum_{\ga} L_{\ba}(k) L_{\ac}^{-1}(k) \right) \phi_{\cb} (k) \right] - \sum_{\ac}  \left[  \left(  \sum_{\gb} L_{\ab}(k) L_{\bc}^{-1}(k) \right)  \phi_{\ca} (k) \right] \\
&= -\sum_{\bc} \gd_{\bc} \phi_{\cb} (k) -  \sum_{\ac} \gd_{\ac}  \phi_{\ca} (k) = -2 \sum_{\ga} \phi_{\ga\ga} (k). \label{eq:FE}
\end{align}
A similar calculation gives the potential enstrophy forcing spectrum $F_G (k)$. We use the operator indicated by Eq. \eqref{eq:Gk}, and we find  that $F_G (k)$ reads:
\begin{align}
F_G (k)  &= \sum_{\abc} L_{\ab} (k) L_{\ac}(k) F_{\bc}(k) \\
 &= \sum_{\abc} L_{\ab} (k) L_{\ac}(k) \sum_\gd  \left[ L_{\bd}^{-1}(k) \phi_{\dc} (k) + L_{\cd}^{-1}(k) \phi_{\db} (k) \right]\\
 &= \sum_{\acd}  L_{\ac}(k) \left[ \sum_\gb L_{\ab} (k) L_{\bd}^{-1}(k) \right] \phi_{\dc} (k) + \sum_{\abd} L_{\ab} (k) \left[ \sum_\gc  L_{\ac}(k) L_{\cd}^{-1}(k) \right] \phi_{\db} (k) \\
 &= \sum_{\acd} \gd_{\ad} L_{\ac}(k) \phi_{\dc} (k) + \sum_{\abd} L_{\ab} (k) \gd_{\ad} \phi_{\db} (k) 
 = 2\sum_{\ab}L_{\ab} (k)  \phi_{\ab} (k). \label{eq:FG}
\end{align}
Note that the $F_E (k)$ calculation is dependent on the symmetry assumption $L_{\ab} (k) = L_{\ba}(k)$, which multi-layer quasi-geostrophic models do satisfy. On the contrary, the  $F_G (k)$ calculation is not dependent on this symmetry assumption. 

Because of the dependence of the streamfunction-forcing spectrum $\phi_{\ga\gb} (k)$ on the streamfunction  $\gy_\ga$, it is not possible to predict the forcing spectra $F_E (k)$ and $F_G (k)$ solely from the statistical properties of the forcing term $f_\ga$. The sole exception is the case where $f_\ga$ is a random Gaussian field that is delta-correlated in time such that
\begin{equation}
\avg{f_\ga (\bfx_1, t_1) f_\gb (\bfx_2, t_2)} = 2Q_{\ga\gb} (\bfx_1, \bfx_2) \gd (t_1-t_2),
\end{equation}
where $Q_{\ga\gb} (\bfx_1, \bfx_2)$  is assumed to be known. Then, it can be shown that the streamfunction-forcing spectrum $\phi_{\ga\gb} (k)$ is given by
\begin{equation}
\phi_{\ab} (k) = \sum_\gc \cQ_{\ac}(k) L^{-1}_{\bc} (k).
\end{equation}
Here $\cQ_{\ac}(k)$ is the correlation spectrum of the forcing term $f_\ga$ given by
\begin{equation}
\cQ_{\ac}(k) = \dD{k} \int \df{\bfx}\df{\bfy}\df{\bfz} \; P(k|\bfx-\bfy) P(k|\bfx-\bfz) Q_{\ac}(\bfy,\bfz).
\end{equation}
We give a detailed derivation of this result in Appendix~\ref{app:StreamfunctionForcingSpectrum}.

\section{Estimating the downscale injection rates}

We will now consider three different arguments for estimating the ratio $\gn/\gee$  of the potential enstrophy injection rate $\gn$ to the energy injection rate $\gee$ to the downscale inertial range. It should be noted here that a careful distinction needs to be made between the total injection rates to the flow, given by integrating Eq.~\eqref{eq:FE} and Eq.~\eqref{eq:FG}, versus the partial injection rates to the coexisting downscale cascades of potential enstrophy and energy. For the purpose of estimating the transition wavenumber $k_t \sim \sqrt{\gn/\gee}$, it is the partial downscale injection rates $\gn$ and $\gee$ that are relevant. These partial downscale injection rates are dependent on both the forcing term and the Ekman term, and that complicates the task ahead. 

In the first argument, we assume that  the model is forced exclusively through the temperature equation, and we disregard the effect of the Ekman term to the downscale injection rates.  In the second argument, we consider the combined effect of thermal forcing and the asymmetric Ekman dissipation term to the energy and potential enstrophy partial downscale injections rates. We will show that the Ekman term tends to shift the transition wavenumber $k_t$ towards large scales, but this claim is predicated on the hypothesis that the potential energy spectrum dominates the kinetic energy spectrum at the forcing range, and the status of this hypothesis is presently uncertain. This prompts us to consider the third argument, where we force both potential vorticity equations with  random delta-correlated in time Gaussian forcing where  the forcing on the bottom layer is suppressed using a control factor $\mu$. The underlying modeling assumption is that the asymmetric Ekman term suppresses forcing on the bottom layer but not at the top layer. 

By combining our results from these three arguments, we will argue that  the two-layer quasi-geostrophic model tends to place the transition wavenumber $k_t$ near the Rossby deformation wavenumber $k_R$. It should be noted that, due to interlayer interaction, it is not obvious whether the Ekman term actually suppresses predominantly the bottom-layer forcing. Furthermore, for models where the forcing is flow-dependent, there is further uncertainty on the effect of the Ekman term on forcing and the overall adjustment of the  partial downscale injection rates. These caveats are discussed further below.

We begin the argument by rewriting our general expressions for the  energy forcing spectrum $F_E (k)$ and  the  potential enstrophy forcing spectrum $F_G (k)$ in terms of the streamfunction-forcing spectrum $\phi_{\ga\gb} (k)$ for the special case of the two-layer model by substituting the corresponding matrix $L_{\ga\gb} (k)$ from Eq.~\eqref{eq:TheMatrixL}. It is easy to see that  the  energy forcing spectrum $F_E (k)$ reads:
\begin{equation}
F_E (k) = -2 \sum_{\ga} \phi_{\ga\ga} (k) = -2[\phi_{11}(k) +\phi_{22}(k)].
\label{eq:FE}
\end{equation}
Likewise,  the  potential enstrophy forcing spectrum $F_G (k)$ reads:
\begin{align}
F_G (k)  &=2\sum_{\ab}L_{\ab} (k)  \phi_{\ab} (k)\\
&= 2[L_{11}(k)  \phi_{11}(k)  + L_{12}(k)  \phi_{12}(k)  + L_{21}(k)  \phi_{21}(k)  + L_{22}(k)  \phi_{22}(k) ]\\
&= -2 (k^2+k_R^2/2)[\phi_{11}(k) +\phi_{22}(k) ] + 2(k_R^2/2)[\phi_{12}(k) +\phi_{21}(k)].
\label{eq:FG}
\end{align}
Using these expressions as our point of departure we now proceed with our analysis of the three energy and potential enstrophy  partial downscale injection rate estimates.

\subsection{Estimate  1: Thermal forcing neglecting Ekman term}

Under thermal forcing we may assume that the potential vorticity equations are forced with $f_1=\gf$ and $f_2=-\gf$. Let us define  $\Phi_1 (k) = \innerf{\gf}{\gy_1}{k}$  and  $\Phi_2 (k) = \innerf{\gf}{\gy_2}{k}$ as the streamfunction correlators with $\gf$. We may therefore write the components of the streamfunction-forcing spectrum  $\phi_{\ga\gb} (k)$ as:
\begin{align}
\phi_{11}(k) &= \innerf{\gf}{\gy_1}{k} = \Phi_1 (k), \\
\phi_{12}(k) &= \innerf{\gf}{\gy_2}{k} = \Phi_2 (k), \\
\phi_{21}(k) &= \innerf{-\gf}{\gy_1}{k} = -\Phi_1 (k), \\
\phi_{22}(k) &= \innerf{-\gf}{\gy_2}{k} = -\Phi_2 (k).
\end{align}
It follows that the energy forcing spectrum  $F_E (k)$ is given by:
\begin{equation}
F_E (k) = -2[\phi_{11}(k) +\phi_{22}(k)] = -2[\Phi_1 (k) - \Phi_2 (k)],
\end{equation}
and the potential enstrophy forcing spectrum $F_G (k)$ is given by:
\begin{align}
F_G (k) &= -2 (k^2+k_R^2/2)[\phi_{11}(k) +\phi_{22}(k) ] + 2(k_R^2/2)[\phi_{12}(k) +\phi_{21}(k) ]\\
 &= -2 (k^2+k_R^2/2)[\Phi_1 (k) - \Phi_2 (k)] + 2(k_R^2/2)[\Phi_2 (k) - \Phi_1 (k)]\\
 &= -2(k^2+k_R^2) [\Phi_1 (k) - \Phi_2 (k)].
\end{align}
We see  that the energy forcing spectrum  $F_E (k)$ and  the potential enstrophy forcing spectrum $F_G (k)$ are related as
\begin{equation}
F_G (k)=  (k^2+k_R^2) F_E (k).
\end{equation}
In the limit $k \sim k_f \ll k_R$ we find that $F_G (k)\sim k_R^2 F_E (k)$, which implies that the ratio $\gn/\gee$ of injected potential  enstrophy rate $\gn$ to injected energy rate $\gee$ is approximately equal to the square of the Rossby deformation wavenumber $k_R$. It follows that, if all of the injected energy and enstrophy cascade downscale and get successfully dissipated at small scales, we will then have a double  potential enstrophy--energy cascade with transition wavenumber $k_t \sim k_R$. We suggest therefore, with some caveats to be discussed further below, that the two-layer quasi-geostrophic model provides a competent explanation for why the Nastrom-Gage spectrum exhibits a transition from $k^{-3}$ scaling to $k^{-5/3}$ scaling near the Rossby deformation wavenumber $k_R$.

It should be stressed that this calculation neglects the effect of Ekman dissipation of the energy and potential enstrophy injection rates, and is only good as a first approximation. In the next two arguments we will suggest that the Ekman term may tend to decrease $k_t$ further and that it is unlikely that it can suppress the partial downscale energy injection rate, as  one typically expects in two-dimensional turbulence. 

\subsection{Estimate  2: Thermal forcing with asymmetric Ekman dissipation}

Let us now consider the effect of asymmetric Ekman dissipation on the  partial downscale rates of  potential enstrophy and energy injection. It is well-known that in two-dimensional turbulence, large-scale dissipation predominantly dissipates most of the injected energy while allowing a considerable fraction of enstrophy to cascade to small scales. This is evidenced by all of the numerical simulations that have successfully reproduced the enstrophy cascade with $k^{-3}$ spectrum \citep{article:Alvelius:2000,article:Falkovich:2002,article:Ishihira:2001}. If the same thing were to occur in the two-layer quasigeostrophic model, it would undermine our previous argument concerning the location of the transition wavenumber $k_t$, because in that argument  we  assumed that most of the injected energy cascades downscale. 

As far as the downscale cascades are concerned, they ``feel'' forcing from both the forcing term and the Ekman term. It is therefore necessary to define the effective forcing spectra $\cF_E (k)$ and $\cF_G (k)$ in which the effects of asymmetric forcing and Ekman dissipation are included together.  We use calligraphic notation  to distinguish them from the forcing spectra  $F_E (k)$ and $F_G (k)$ defined via Eq.~\eqref{eq:FE} and Eq.~\eqref{eq:FG}. The partial injection rates $\gn$ and $\gee$ to the downscale cascades are given by integrating $\cF_E (k)$ and $\cF_G (k)$. As a matter of mathematical expediency, we can still calculate  $\cF_E (k)$ and $\cF_G (k)$ via  Eq.~\eqref{eq:FE} and Eq.~\eqref{eq:FG} by \emph{redefining} the forcing terms $f_1$ and $f_2$ to include the asymmetric Ekman term. The calculation below shows that the effect of the asymmetric Ekman term is to tend to decrease the effective energy forcing spectrum  $\cF_E (k)$, thereby  acting as an energy sink. However, as long as the potential energy spectrum dominates the kinetic energy spectrum at the forcing range, the effective potential enstrophy forcing spectrum $\cF_G (k)$ will decrease much faster, thereby shifting the transition wavenumber to smaller scales. It should be stressed that in most quasi-geostrophic models, the forcing term is flow-dependent, so the reaction of the flow to the effect of the asymmetric Ekman term adds another degree of uncertainty to the above claims. On the other hand, for random Gaussian forcing that is delta-correlated in time, the reaction of the flow does not affect the effective forcing spectra.

Let us now elaborate on the above argument. We begin by \emph{redefining} $f_1=\gf$ and $f_2=-\gf-\nu_E \del^2 \gy_2$, as discussed above. Recall that $C_{12}(k) = \innerf{\gy_1}{\gy_2}{k}$, and let us define $U_1 (k) = \innerf{\gy_1}{\gy_1}{k}$ and $U_2 (k) = \innerf{\gy_2}{\gy_2}{k}$. It follows that the streamfunction-forcing spectra $\phi_{\ga\gb} (k)$  are given by:
\begin{align}
\phi_{11}(k) &= \innerf{\gf}{\gy_1}{k} = \Phi_1 (k), \\
\phi_{12}(k) &= \innerf{\gf}{\gy_2}{k} = \Phi_2 (k), \\
\phi_{21}(k) &= \innerf{-\gf-\nu_E \del^2 \gy_2}{\gy_1}{k} = -\Phi_1 (k) + \nu_E k^2 C_{12}(k),\\
\phi_{22}(k) &= \innerf{-\gf-\nu_E \del^2 \gy_2}{\gy_2}{k} = -\Phi_2 (k) + \nu_E k^2 U_2 (k).
\end{align}
Substituting to Eq.~\eqref{eq:FE}, we find that the effective energy forcing spectrum  $\cF_E (k)$ is given by:
\begin{align}
\cF_E (k) &= -2[\Phi_1 (k) - \Phi_2 (k) + \nu_E k^2 U_2 (k)] \\
&= F_E (k)-2\nu_E k^2 U_2 (k).
\end{align}
 Since $U_2 (k)$  is positive-definite, we have $U_2 (k)\geq 0$, and therefore the asymmetric Ekman dissipation term decreases the rate of energy injection. Note that if the forcing term $\gf$   is dependent on the flow, as is the case in both the Tung-Orlando and Tulloch-Smith simulations \citep{article:Orlando:2003,article:Smith:2009}, the decrease by the $\nu_E k^2 U_2 (k)$  term could be counteracted by an increase by the $2[\Phi_1 (k) - \Phi_2 (k)]$  term, as pointed out to the author by an anonymous referee. However, if $\gf$   is given as a standard independent random forcing term, which is common practice in turbulence simulations, then $\cF_E (k)$  is  decreased by the Ekman term.

Similarly, the  effective  potential enstrophy forcing spectrum  $\cF_G (k)$ is given by 
\begin{align}
\cF_G (k)- (k^2+k_R^2) \cF_E (k) 
 &= k_R^2 [\phi_{11}(k)+\phi_{12}(k)+\phi_{21}(k)+\phi_{22}(k)]  \label{eq:FGminFEformula}\\
&= \nu_E k_R^2 k^2 [C_{12}(k) +U_2(k) ],
\end{align}
and it follows that $\cF_G (k)$  is given by
\begin{equation}
\cF_G (k) = (k^2+k_R^2) \cF_E (k)+\nu_E k_R^2 k^2 [C_{12}(k) +U_2(k) ].\label{eq:FGwithEkman}
\end{equation}
We see that whether the Ekman term tends to shift $k_t$ upscale or downscale  depends on the sign of $C_{12}(k) +U_2(k)$. It is already known that $U_2(k)\geq 0$. However, $C_{12}(k)$  can be either positive or negative. The condition $C_{12}(k) < 0$  is necessary but not sufficient in ensuring a transition wavenumber shift towards large scales. The necessary and sufficient condition for effecting such a shift is $U_2 (k)+C_{12}(k) < 0$. On the other hand, the condition $C_{12} (k) < U_2 (k)$ is sufficient to ensure that the Ekman term dissipates potential enstrophy, which is expected on physical grounds. To show this, we rewrite the effective potential enstrophy forcing spectrum $\cF_G (k)$ in terms of the potential enstrophy forcing spectrum $F_G (k)$ as follows: 
\begin{align}
\cF_G (k) &= (k^2+k_R^2) [F_E (k) -  2\nu_E k^2 U_2(k)] +\nu_E k_R^2 k^2 [C_{12}(k) +U_2(k) ]\\
&= F_G (k) + \nu_E k^2 [k_R^2 (C_{12} (k)- U_2 (k)) - 2k^2 U_2 (k)].
\end{align}
Consequently, the Ekman term dissipates potential enstrophy if $k_R^2 (C_{12} (k)- U_2 (k)) - 2k^2 U_2 (k)<0$   for all wavenumbers $k$ in the forcing range. Since $U_2 (k) > 0$, due to being positive-definite, it is sufficient that $C_{12} (k) < U_2 (k)$.

We can gain some insight on $C_{12}(k)$ by relating it with the kinetic and potential energy spectra $E_K (k)$ and $E_P (k)$  which are defined as follows: Let $\gy \equiv (\gy_1 + \gy_2)/2$ and $\gt \equiv (\gy_1 - \gy_2)/2$. So, $\gy_1 = \gy+\gt$ and $\gy_2 = \gy-\gt$. Following \citet{article:Salmon:1978,article:Salmon:1980}, the definitions of the spectra $E_K (k)$, $E_P (k)$, and $E_C (k)$ in terms of $\gy$ and $\gt$ are given by:
\begin{align}
E_K (k) &= 2k^2 \innerf{\gy}{\gy}{k}, \\
E_P (k) &= 2(k^2 + k_R^2) \innerf{\gt}{\gt}{k}, \\
E_C (k) &= 2k^2 \innerf{\gy}{\gt}{k}.
\end{align}
 It can be shown that the definitions are self-consistent, i.e. $E (k)=E_K (k)+E_P (k)$. It is easy now to write $C_{12}(k)$ in terms of $E_K (k)$ and $E_P (k)$:
\begin{align}
C_{12}(k) &= \innerf{\gy_1}{\gy_2}{k} = \innerf{\gy+\gt}{\gy-\gt}{k} =  \innerf{\gy}{\gy}{k} - \innerf{\gy}{\gt}{k} + \innerf{\gt}{\gy}{k} - \innerf{\gt}{\gt}{k}\\
&= \innerf{\gy}{\gy}{k} - \innerf{\gt}{\gt}{k} = \frac{E_K (k)}{2k^2} - \frac{E_P (k)}{2(k^2+k_R^2)}.
\end{align}
We see that requiring $E_K (k)\ll E_P (k)$ for all wavenumbers $k$ in the forcing range is sufficient to ensure that  $C_{12}(k)$ be negative. 

To obtain a necessary and sufficient condition, we first note that 
\begin{align}
U_2(k) &= \innerf{\gy_2}{\gy_2}{k} = \innerf{\gy-\gt}{\gy-\gt}{k} = \innerf{\gy}{\gy}{k} - 2 \innerf{\gy}{\gt}{k} + \innerf{\gt}{\gt}{k}\\
&=  \frac{E_K (k)}{2k^2} + \frac{E_P (k)}{2(k^2+k_R^2)} - \frac{E_C (k)}{k^2}.
\end{align}
It follows that 
\begin{equation}
U_2(k) + C_{12}(k) = \frac{E_K (k)-E_C (k)}{k^2},
\end{equation}
and therefore $U_2(k) + C_{12}(k) < 0$ if and only if $E_K (k)<E_C (k)$ for all wavenumbers $k$ in the forcing range. It should be stressed that, as far as the transition wavenumber $k_t$ is concerned, the relevant requirement is that the ratio $\cF_G (k)/\cF_E (k)$  should be decreased by the Ekman term. It is easy to see from Eq.\eqref{eq:FGwithEkman} that $\cF_G (k)/\cF_E (k)$ is a linear function of $\nu_E$ with slope $k_R^2 k^2 [C_{12}(k) +U_2(k) ]$. Thus, the condition $E_K (k)<E_C (k)$ is indeed the needed necessary and sufficient condition.

Without a detailed phenomenological understanding of the two-layer model, it is hard to say whether this condition is satisfied. \citet{article:Salmon:1980} has argued that in the two-layer model, energy is being injected as potential energy and gets converted to kinetic energy near the Rossby wavenumber $k_R$. We may therefore expect the potential energy to remain dominant in the forcing range, provided that most kinetic energy does not inversely cascade back to large scales again. Nevertheless, this is an open question that should be carefully investigated via numerical simulations. In the next section we will provide an alternate argument supporting the claim of a  transition wavenumber shift to large scales, predicated on the hypothesis that the Ekman dissipation term suppresses forcing only at the lower-layer, thereby assuming that the interlayer interaction does not propagate Ekman  dissipation into the top layer. The spectrum $C_{12}(k)$ captures, in effect, an aspect of the dynamics of this interlayer interaction.

Be that as it may, we note that it is also possible to formulate arguments that suggest the opposite conclusion as follows: As \citet{article:Orlando:2003:1} have shown, an equipartition of kinetic and potential energy is expected in the  extreme baroclinic limit represented by the SQG model (i.e. $E_P (k)/E_K (k) = 1$). The opposite  limit,  if generalized for all scales, is the case of three-dimensional stratified turbulence where \citet{article:Lindborg:2006} observed a $1/3$ distribution of the total energy between potential and kinetic such that $E_P (k)/E_K (k) \sim 1/3$, with the exact ratio being somewhat dependent on the rotation rate. For both cases we have $E_K (k)\geq E_P (k)$. Furthermore, in a recent direct numerical simulation of the full quasi-geostrophic model by \citet{article:Lindborg:2010}, it has been confirmed that the total energy spectrum $E(k)$ is equipartitioned between potential energy $E_P (k)$  and the two horizontal components of kinetic energy, leading to an approximate ratio $E_P (k)/E_K (k) \sim 1/2$, consistent with the theory of \citet{article:Charney:1971}. A deviation seems to occur in the forcing range where $E_P (k)/E_K (k) \sim 1$, but any value between $1/2$  and $1$ violates the sufficient condition  $E_K (k)\ll E_P (k)$. It should be noted, however, that the simulation of \citet{article:Lindborg:2010} is forced symmetrically instead of antisymmetrically, and it is uncertain how that may affect the partition ratio of energy between kinetic energy and potential energy. Also uncertain is the effect of restricting the full quasi-geostrophic model to two layers and using asymmetric instead of symmetric dissipation.

At this point, one could argue that if the Ekman term in $\cF_G (k)$  is negligible, then it doesn't matter either way whether $\cF_G (k)$ is increasing or decreasing. We will now argue, using a phenomenological order of magnitude estimate, that the Ekman adjustment of the potential enstrophy forcing spectrum $\cF_G (k)$ is not expected to be negligible. The argument is as follows: On the assumption that most injected energy cascades to small scales, $\cF_E (k)$ is proportional to the downscale energy flux $\gee$. If we also assume that the forcing spectrum is spread over a wavenumber interval with width proportional to the average forcing wavenumber $k_f$, then we get the dimensional  estimate $\cF_E (k) \sim \gee/k_f$. This estimate is a lower bound for $\cF_E (k)$  since, as an anonymous referee noted, it is possible, in principle, for the forcing spectrum to be concentrated on a peak with width $\gD k$ narrower than $k_f$. We also assume that $k^2 [C_{12}(k) +U_2(k) ]$, which has the dimension of the energy spectrum, scales as $k^2 [C_{12}(k) +U_2(k) ] \sim \gn^{2/3} k_f^{-3}$, consistent with the downscale potential enstrophy cascade spectrum. Putting these two phenomenological estimates together, for forcing-range wavenumbers $k \sim k_f \ll k_R$, we estimate the two terms on the right-hand-side of Eq.~\eqref{eq:FGwithEkman} as: 
\begin{align}
\cA &\equiv (k^2+k_R^2) \cF_E (k) \sim \gee k_R^2/k_f,\\
\cB &\equiv \nu_E k_R^2 k^2 (C_{12} (k) + U_2 (k)) \sim \nu_E k_R^2 \gn^{2/3} k_f^{-3}.
\end{align}
Using the relation $ \gn\sim \gee k_t^2$  between the potential enstrophy flux $ \gn$  and the energy flux $\gee$, we find that the ratio of the two terms is estimated by: 
\begin{align}
\frac{\cA}{\cB} &\sim \frac{\gee k_R^2 k_f^{-1}}{\nu_E k_R^2 \gn^{2/3} k_f^{-3}}
\sim \frac{\gee k_f^{2}}{\nu_E \gn^{2/3}} 
\sim \frac{ \gn k_f^{2}}{\nu_E \gn^{2/3} k_t^2}
\sim \frac{\gn^{1/3}}{\nu_E}\fracp{k_f}{k_t}^2.
\end{align}
For the potential enstrophy flux $\gn$  we use the value $\gn\sim 10^{-15} \text{s}^{-3}$  estimated by \citet{article:Lindborg:2001} by structure function analysis. For the Ekman coefficient $\nu_E$, we use the number $\nu_E \sim (6.7 \text{days})^{-1} \sim 10^{-6} \text{s}^{-1}$  by \citet{article:Orlando:2003}. Finally, from the Nastrom-Gage spectrum itself, we can estimate $k_t \sim 10^{-3} \text{km}^{-1}$  and $k_f \sim 10^{-4} \text{km}^{-1}$   for the transition and forcing wavenumbers. Using these numbers, we find that $\cA/\cB \sim 10^{-1}$, which implies that the terms $\cA$ and $\cB$ are comparable within one order of magnitude, so the effect of the constant coefficients is likely to play an important role in deciding which term is dominant. Note that if the forcing spectrum is concentrated on a peak with width  $\gD k$ with $\gD k \ll k_f$, then $\cF_E (k)$  is increased, thus the change to the ratio $\cA/\cB$ in turn indicates  a diminishing impact of the Ekman term on the transition wavenumber $k_t$. Consequently, within the framework of the above phenomenology, our estimate of the  $\cA/\cB$  ratio represents a worst-case scenario, in the sense that the effect of the Ekman term can't be stronger than this estimate. 

In light of the above, it is  very important to further investigate, with numerical simulations, the effect of the Ekman term on the injection rates, using both the quasi-geostrophic model of \citet{article:Orlando:2003} and the quasi-geostrophic model of \citet{article:Smith:2009}. Specifically, for the case of the two-layer model, future numerical studies should, at the very least, investigate the interlayer spectrum $C_{12}(k)$ and the partition of energy between kinetic energy and potential energy.

\subsection{Estimate 3: Asymmetric random forcing}

In the previous case, we have seen that the effect of asymmetric Ekman dissipation on the forcing range is to tend to decrease the rate of energy injection. There is, however, ambiguity regarding whether the enstrophy injection rate is increasing or decreasing, and whether the Ekman term shifts the transition wavenumber $k_t$ towards small scales or large scales. The underlying problem is that,  due to the effect of the layer to layer interaction on the relationship between potential vorticity and streamfunction, it is not obvious whether dissipating the bottom layer streamfunction $\gy_2$ is equivalent to dissipating the bottom-layer potential vorticity $q_2$. On the other hand, we will show now that if the bottom-layer forcing is directly suppressed via a control factor $\mu$, that will indeed result in a reduction of the ratio $F_G (k)/F_E (k)$  in the forcing range.

To that end, let us assume that the forcing terms for the top and bottom layers respectively are  $f_1=\gf$ and $f_2=-\mu \gf$  with $0 < \mu < 1$. Decreasing $\mu$ increases the suppression of the bottom-layer forcing term $f_2$. We also assume that $\gf$ is a delta-correlated in time random Gaussian field with correlation spectrum $\cQ(k)$. In appendix \ref{app:RandomGaussianForcing}, we show that the forcing-streamfunction spectra $\phi_{\ga\gb} (k)$ can be expressed in terms of $\cQ(k)$ as follows:
\begin{equation}
\gf_{\ab} (k) = \frac{-\cQ (k) \gy_{\ab} (k)}{2 k^2 (k^2+k_R^2)}.
\end{equation}
Here, $\gy_{\ab}$ are given by:
\begin{align}
\gy_{11} (k)  &= (2k^2+k_R^2)-\mu k_R^2, \\
\gy_{12} (k)  &= k_R^2-\mu (2k^2+k_R^2), \\
\gy_{21} (k)  &= -\mu (2 k^2+k_R^2) + \mu^2 k_R^2, \\
\gy_{22} (k)  &= -\mu k_R^2 + \mu^2 (2k^2+k_R^2).
\end{align}
Without explicitly calculating the forcing spectra $F_E (k)$ and $F_G (k)$, we can readily argue that since 
\begin{align}
\gy_{11} (k) + \gy_{12} (k) + \gy_{21} (k) + \gy_{22} (k)  
&= 2(k^2+k_R^2)(1-\mu)^2,
\end{align}
 it follows that 
\begin{align}
F_G (k) - (k^2+k_R^2) F_E (k) &= k_R^2 [\phi_{11}(k)+\phi_{12}(k)+\phi_{21}(k)+\phi_{22}(k)]\\
&= \frac{-\cQ (k) [\gy_{11} (k) + \gy_{12} (k) + \gy_{21} (k) + \gy_{22} (k)]}{2 k^2 (k^2+k_R^2)}\\
&= \frac{-\cQ (k) [2(k^2+k_R^2)(1-\mu)^2]}{2 k^2 (k^2+k_R^2)} =  - \frac{k_R^2 (1-\mu)^2 \cQ (k)}{k^2},
\end{align}
via Eq.~\eqref{eq:FGminFEformula} and therefore 
\begin{equation}
F_G (k) =  (k^2+k_R^2) F_E (k)- \frac{k_R^2 (1-\mu)^2 \cQ (k)}{k^2}.
\end{equation}
We note that since the third term in the equation above is always negative,  suppressing the lower-level forcing leads to a large-scale shift of the transition wavenumber $k_t$. For $\mu=1$, as expected, we recover the previously derived relation $F_G (k) =  (k^2+k_R^2) F_E (k)$.

Another way of looking at the problem is by explicitly calculating the ratio $F_G (k)/F_E (k)$ and showing that it decreases with decreasing $\mu=1$. 
As shown in appendix \ref{app:RandomGaussianForcing}, an explicit calculation of the forcing spectra $F_E (k)$ and $F_G (k)$ gives: 
\begin{align}
F_E (k) &= \frac{2\cQ(k)[2(1+\mu^2) k^2 + (1-\mu)^2 k_R^2]}{2k^2 (k^2+k_R^2)}, \\
F_G (k) &= 2\cQ(k) (1+\mu^2).
\end{align}

 For the antisymmetric case $\mu=1$, the energy forcing spectrum  $F_E (k)$  reduces to $F_E (k) = 4\cQ(k)/(k^2+k_R^2)$  and the enstrophy forcing spectrum  $F_G (k)$ reduces to $F_G (k) = 4\cQ(k)$, thereby recovering our previous more generally applicable result $F_G (k)=  (k^2+k_R^2) F_E (k)$, which suggests a transition wavenumber $k_t\sim k_R$, in the limit $k \sim k_f \ll k_R$.

For the extreme case $\mu=0$, whereby the bottom-layer forcing is completely suppressed, again under the limit $k  \sim k_f \ll k_R$, the energy forcing spectrum $F_E (k)$  is given by 
\begin{equation}
F_E (k) = \frac{2\cQ(k)[2k^2+k_R^2]}{2k^2 (k^2+k_R^2)}  \sim \frac{2\cQ(k)k_R^2}{2k^2 k_R^2}  \sim \frac{\cQ(k)}{k^2},
\end{equation} 
and the enstrophy forcing spectrum  $F_G (k)$  is given by $F_G (k) = 2Q(k)$. It follows that $F_G (k) \sim 2k^2 F_E (k)$,  which suggests a reduced transition wavenumber $k_t\sim 2k_f$.

The two extreme cases  $\mu=1$  and  $\mu=0$ indicate that the ratio $F_G (k)/F_E (k)$  decreases with decreasing  $\mu$ from approximately  $k_R^2$  to $2k_f^2$. We can confirm that this is indeed the case by taking  the partial derivative with respect to the parameter  $\mu$. The partial derivative reads
\begin{align}
\pD{\mu}\fracb{F_G (k)}{F_E (k)} &= \pD{\mu}\fracb{2(1+\mu^2)\cQ(k)2k^2 (k^2+k_R^2)}{2\cQ(k)[2(1+\mu^2) k^2 + (1-\mu)^2 k_R^2]}\\
&= 2k^2 (k^2+k_R^2)\pD{\mu}\fracb{1+\mu^2}{[2(1+\mu^2) k^2 + (1-\mu)^2 k_R^2]}\\
&=  2k^2 (k^2+k_R^2)\fracb{2k_R^2 (1-\mu)(1+\mu)}{[2(1+\mu^2) k^2 + (1-\mu)^2 k_R^2]^2}.
\end{align}
 For $0 < \mu < 1$, it is easy to see that every factor is positive, and therefore 
\begin{equation}
\pD{\mu}\fracb{F_G (k)}{F_E (k)} > 0.
\end{equation} 
Consequently, the ratio $F_G (k)/F_E (k)$ decreases with decreasing $\mu$. We conclude that if the asymmetric Ekman damping term on the bottom-layer streamfunction $\gy_2$ indeed suppresses the effective forcing of the bottom-layer potential vorticity, then the ratio $F_G (k)/F_E (k)$ will tend to decrease, thereby indicating a tendency to reduce the transition wavenumber $k_t$.

\section{Conclusions and Discussion}

In the present paper, we have sought out to explain why  the transition from $k^{-3}$ scaling to  $k^{-5/3}$ scaling in the Nastrom-Gage  spectrum occurs near the Rossby deformation wavenumber $k_R$, where the atmospheric turbulence is still governed under quasi-geostrophic dynamics instead of three-dimensional dynamics. According to the Tung-Orlando theory \citep{article:Orlando:2003}, the entire Nastrom-Gage spectrum represents a downscale potential enstrophy cascade that co-exists with a downscale energy cascade. The location of the transition wavenumber $k_t$ is thereby controlled by the ratio $\gn/\gee$  of the downscale potential enstrophy flux $\gn$ over the downscale energy flux $\gee$ and given by $k_t \sim \sqrt{\gn/\gee}$. That ratio, in turn, depends on the large-scale forcing and the effect of large-scale dissipation on the injection of potential enstrophy and energy.

We have shown that in the two-layer quasi-geostrophic model, which is reasonably applicable in the forcing scales, thermal forcing leads to antisymmetric forcing of the potential vorticity layer equations. This, in turn,  yields a ratio $\gn/\gee$  of the potential enstrophy injection rate  $\gn$ over the energy injection rate  $\gee$ that is approximately equal to $k_R^2$. So, if most of the injected potential enstrophy and energy cascades towards small scales, then the transition wavenumber $k_t$ will be approximately equal to $k_R$. 

At this point, one might object by arguing, drawing from an analogy with two-dimensional turbulence, that the large-scale Ekman dissipation will get rid of  most of the injected energy at the forcing range while allowing a considerable amount of potential enstrophy to cascade to small scales. As it turns out, it is far from obvious that the two-layer model behaves in this manner. In general, the Ekman term always dissipates some amount of energy, and may or may not dissipate potential enstrophy, depending on the sign and magnitude of the interlayer spectrum $C_{12} (k)$. We have shown that if the potential energy spectrum $E_P (k)$ dominates the kinetic energy spectrum $E_K  (k)$ in the forcing range, then the downscale potential enstrophy injection rate $\gn$ will be dampened faster than the downscale energy injection rate $\gee$.  The resulting reduction in the $\gn/\gee$ ratio will tend to shift the transition wavenumber $k_t$ towards large scales. This tendency becomes  exact,  if the forcing used in these simulations is made independent of the flow.  

Unfortunately, there is some ambiguity in the results of our direct analysis of the Ekman term, due to the dependence of the direction of the transition wavenumber shift on the spectral distribution of the energy between kinetic and potential energy.  Using a random Gaussian forcing model, we have shown that, under the assumption that the Ekman term suppresses forcing predominantly at the bottom layer,  the ratio $\gn/\gee$ will be decreased, thereby shifting the transition wavenumber $k_t$ to larger scales.  While this assumption may seem obvious, on physical grounds, it requires us to disregard the possibility of Ekman dissipation being propagated to the top layer via the nonlinear interlayer interaction.  Without a more detailed understanding of the phenomenology of the two-layer model, and especially the interlayer spectrum $C_{12}(k)$, this is as far as we can go on this problem in terms of theory.

Another problem with our argument is that it is only one-half of the whole story. In order for the injected potential enstrophy and energy to form a steady-state cascade, it is also necessary that the small-scale dissipation terms be able to dissipate the potential enstrophy and energy at the same rate with which they are injected. In a strictly two-dimensional model, this is impossible, because the potential enstrophy and energy fluxes $\Pi_G (k)$  and $\Pi_E (k)$  are constrained by the inequality $k^2 \Pi_E (k) - \Pi_G (k)<0$,  for all wavenumbers $k$ not in the forcing range \citep{article:Tung:2005,article:Tung:2005:1}. However, as we have shown previously in \citet{article:Tung:2007:1}, the asymmetric Ekman  dissipation  term can potentially cause this flux inequality to be violated. If that occurs, then a  transition from $k^{-3}$ to  $k^{-5/3}$ scaling is possible near the wavenumber $k_t$ where the aforementioned flux inequality breaks down. Unfortunately, it is not easy to derive a rigorous necessary and sufficient condition for violating the flux inequality, in the form of a lower bound for $\nu_E$,  without introducing phenomenological assumptions. In light of the controversy with the Tung-Orlando simulation \citep{article:Smith:2004,article:Tung:2004,article:Tung:2007}, this energy dissipation sink problem remains an open question. On the other hand, we are  quite certain that this flux inequality was successfully violated in the more sophisticated two-mode two-layer quasi-geostrophic model of \citet{article:Smith:2009}, which produced coexisting cascades of potential enstrophy and energy consistent with the Tung-Orlando theory. We do not yet have a detailed mathematical understanding of how this violation came about.



Ultimately, the question of whether QG models can break the flux inequality is somewhat academic, albeit interesting. As \citet{article:Lindborg:2007} has shown, at scales less than $100$km, the assumptions that underlie the quasi-geostrophic model break down. This breakdown  acts in our favor by giving us an anomalous energy dissipation sink at large $k$, thereby further facilitating the breakdown of the flux inequality. What is less obvious is whether there is still an effective potential enstrophy dissipation sink at small scales, occurring either at length scales where the flow is still stratified or via a violation of potential enstrophy conservation at even smaller scales where the flow becomes entirely three-dimensional.
 If yes, then we have a full accounting of the entire process: quasi-geostrophic dynamics is thus responsible for injecting potential enstrophy and energy at a proportion leading to $k_t \sim k_R$, and three-dimensional dynamics is responsible for dissipating both at small scales. If no, then the widely accepted interpretation of the  $k^{-3}$ part of the Nastrom-Gage spectrum as a downscale potential enstrophy cascade is itself in jeopardy, regardless of whether or not one agrees with all other aspects of the Tung-Orlando theory.  An alternate explanation  of the Nastrom-Gage spectrum as a downscale helicity cascade (with $k^{-7/3}$ scaling instead of $k^{-3}$) coexisting with a downscale energy cascade is the only remaining hypothesis on the table, if we were to completely rule out quasi-geostrophic dynamics for all length scales \citep{article:Tsinober:1993,article:Chkhetiani:1996,book:Moiseev:1999,article:Golbraikh:2006}.

It is fair to say that this paper does not resolve all of the outstanding controversies with respect to the Nastrom-Gage spectrum. For example, we have not yet completely resolved the energy dissipation sink issue in the Tung-Orlando simulation, or the question of whether the $k^{-3}$ part of the Nastrom-Gage spectrum is a helicity cascade  or a potential enstrophy cascade. In spite of extensive numerical evidence, e.g. by \citet{article:Mahlman:1999}, \citet{article:Hamilton:2001}, \citet{article:Skamarock:2004}, \citet{article:Ohfuchi:2006}, and \citet{article:Ohfuchi:2008}, I believe that both  questions are still open at the present time. Furthermore, within the framework of the theory presented in this paper, we have posed the new open question of the effect of Ekman dissipation on shifting the transition wavenumber away from the Rossby wavenumber $k_R$. Underlying all this, is the theoretical question of whether the location of the transition wavenumber $k_t$ is \emph{flexible} and controlled via the magnitude of the two fluxes associated with two independent coexisting cascades, as proposed by \citet{article:Orlando:2003}, or whether it is \emph{inflexible} and pinned down near the Rossby wavenumber $k_R$ by a scaling transition inherent in the nonlinearity, as typified by the Tulloch-Smith SQG model \citep{article:Smith:2006}. While we are advocating for the flexible placement of the transition wavenumber $k_t$, it is fair to say that the question deserves further scrutiny.

An anonymous referee has also raised the question of whether gravity waves can play a role in  the Nastrom-Gage spectrum, as was conjectured by \citet{article:Dewan:1979} and \citet{article:VanZadt:1982}. It is well-known that gravity waves vanish in the quasi-geostrophic limit, therefore they are not expected to be relevant over the quasi-geostrophic range of length scales, as were rigorously determined by \citet{article:Lindborg:2007}. According to \citet{article:Nastrom:1986}, the agreement between the measured wavenumber spectra and  frequency spectra, suggests that the spectrum arises from strong turbulence and not from gravity waves. \citet{article:Nastrom:1985} also noted that ``the energy levels and shapes of the horizontal and vertical energy spectra are not consistent with existing models of internal wave spectra'', with the caveat that the inconsistency could be originating from shortcomings of these internal wave spectral models. Given  these arguments against the gravity wave interpretation of the Nastrom-Gage spectrum, and the ``folklore'' belief that quasi-geostrophic dynamics does not allow a downscale energy cascade, it was necessary for \citet{article:Orlando:2003} to demonstrate that the entire Nastrom-Gage spectrum can be reproduced entirely by quasi-geostrophic dynamics in order to bolster their hypothesis of coexisting cascades of potential enstrophy and energy, even under very restricted two-dimensional approximations of quasi-geostrophic dynamics.  \citet{article:Orlando:2003} however did acknowledge that gravity waves could play a role in enabling the coexistence of the two downscale cascades. 

As for the gravity wave interpretation, many relevant questions are still not settled. For instance, \citet{article:Zagar:2011} point towards a very interesting possibility: Using reprocessed observational data provided by the Japan Meteorological Agency, they showed that after decomposing the total energy into a quasi-geostrophic component and a gravity waves component,  the quasi-geostrophic component yields a $k^{-3}$ potential enstrophy cascade contribution spanning the entire range of resolved length scales, and the gravity wave component yields a  $k^{-5/3}$ energy cascade contribution coexisting over the same range of scales. The total energy spectrum is thus the linear superposition of the two contributions. 

This picture is consistent with the Tung-Orlando theory and the linear superposition hypothesis proposed by \citet{article:Tung:2005,article:Tung:2005:1} and \citet{article:Tung:2006}. As explained by \citet{article:Tung:2006}, the underlying principles involved are universal and originate from the linearity of the underlying statistical mechanics, so we expect them to remain valid, beyond two-dimensional turbulence, in all related dynamical systems that allow the coexistence of cascades of energy and  enstrophy.  Under the scenario of coexisting quasi-geostrophic and gravity wave dynamics, indicated by \citet{article:Zagar:2011}, the transition wavenumber is still entirely controlled by the injection rate ratio $\gn/\gee$, given the confirmed validity of the linear superposition principle. The remaining open question is whether the main results of this paper concerning the injection rates ratio (i.e. $\gn/\gee \sim k_R^2$) can be generalized even beyond quasi-geostrophic models. We believe that further research is needed in that direction. 

\begin{acknowledgements}
It is a pleasure to thank Ka-Kit Tung and Joe Tribbia for discussion and correspondence. The idea of an energy-enstrophy flux inequality was originally communicated to me in e-mail correspondence with Sergey Danilov. All anonymous referees also provided very valuable feedback that went a long way into improving the paper.
\end{acknowledgements}

\appendix

\section{The potential vorticity--streamfunction equations}
\label{app:VorticityStreamfunctionEquations}

In this appendix, we derive the potential vorticity equations Eq.~\eqref{eq:PotVortOne} and Eq.~\eqref{eq:PotVortTwo}  from the relative vorticity equations Eq.~\eqref{eq:RelativeVortOne} and Eq.~\eqref{eq:RelativeVortTwo}  and the mid-layer temperature equation Eq.~\eqref{eq:Temperature}.  Our goal is to demonstrate that the potential vorticity equations are forced anti-symmetrically, a key property for the argument of the present paper, and that the dissipation terms in the  relative vorticity equations retain the same form in the potential vorticity equations. The derivation is dependent on the following properties of the Jacobian $J(a,b)$: 
\begin{align}
J&(a, b+c) = J(a, b) + J(a, c),\\
J&(a+b, c) = J(a, c) + J(b, c),\\
J&(a, a) = 0 \text{ and }  J(a, b) = -J(b, a),\\
J&(\gl a,\mu b) = \gl\mu  J(a, b),
\end{align}
where $\gl$ and $\mu$ are constants. 

The first step is to solve for the vertical velocity $\gw$ in the temperature equation Eq.\eqref{eq:Temperature}. From the geostrophic constraint $T = (2/h)(\gy_1-\gy_2)$ we write the advection term in the temperature equation as:
\begin{align}
J\left( \frac{\gy_1+\gy_2}{2}, T \right) &= \frac{1}{h}[J(\gy_1, \gy_1) - J(\gy_1, \gy_2) + J(\gy_2, \gy_1) - J(\gy_2, \gy_2) ]\\
&=  -\frac{2}{h}J(\gy_1, \gy_2).
\end{align}
It follows that the vertical velocity  $\gw$  reads:
\begin{align}
\gw &= -\frac{f}{N^2}\left[ \pderiv{T}{t} + J\left( \frac{\gy_1+\gy_2}{2}, T \right) - Q_0 \right]\\
&= -\frac{2f}{hN^2}\left[\pD{t} (\gy_1-\gy_2) - J(\gy_1, \gy_2) - \frac{hQ_0}{2} \right],
\end{align}
and therefore 
\begin{equation}
\frac{2f}{h}\gw = -\frac{k_R^2}{2}\left[\pD{t} (\gy_1-\gy_2) - J(\gy_1, \gy_2) - \frac{hQ_0}{2} \right].
\end{equation}
Here we have defined the Rossby deformation wavenumber $k_R = 2\sqrt{2} f/(hN)$.

The next step is to define the potential vorticities $q_1$ and $q_2$ for the top and bottom layers correspondingly as: 
\begin{align}
q_1 &=  \del^2 \gy_1 + f + \frac{k_R^2}{2}(\gy_2-\gy_1), \\
q_2 &=  \del^2 \gy_2 + f - \frac{k_R^2}{2}(\gy_2-\gy_1). 
\end{align}
The advection terms  $J(\gy_1, q_1)$ and $J(\gy_2, q_2)$ of the potential vorticities with respect to the streamfunctions  $\gy_1$ and $\gy_2$ are given by:
\begin{align}
J(\gy_1, q_1) &= J(\gy_1, \gz_1+f+(k_R^2/2)(\gy_2-\gy_1)) \\
&= J(\gy_1, \gz_1+f) + \frac{k_R^2}{2} J(\gy_1,\gy_2),
\end{align}
and
\begin{align}
J(\gy_2, q_2) &= J(\gy_2, \gz_2+f-(k_R^2/2)(\gy_2-\gy_1)) \\
&= J(\gy_1, \gz_1+f) + \frac{k_R^2}{2} J(\gy_2,\gy_1)\\
&= J(\gy_1, \gz_1+f) - \frac{k_R^2}{2} J(\gy_1,\gy_2).
\end{align}
Therefore, differentiating the top-layer potential vorticity  $q_1$  with respect to the time $t$  gives:
\begin{align}
\pderiv{q_1}{t} &= \pderiv{\gz_1}{t} + \frac{k_R^2}{2}\pD{t}(\gy_2-\gy_1) \\
&= -J(\gy_1, \gz_1+f) - \frac{2f}{h}\gw + d_1 + \frac{k_R^2}{2}\pD{t}(\gy_2-\gy_1)\\
&= -J(\gy_1, \gz_1+f) + \frac{k_R^2}{2}\left[\pD{t} (\gy_1-\gy_2) - J(\gy_1, \gy_2) - \frac{hQ_0}{2} \right] + d_1 + \frac{k_R^2}{2}\pD{t}(\gy_2-\gy_1)\\
&= -J(\gy_1, \gz_1+f) - \frac{k_R^2}{2} J(\gy_1, \gy_2) - \frac{hk_R^2}{4}Q_0 + d_1\\
&= -J(\gy_1, q_1) - Q + d_1,
\end{align}
with $Q$ defined as $Q= hk_R^2 Q_0/4$. Likewise, differentiating the bottom-layer potential vorticity $q_2$ with respect to the time $t$ gives:
\begin{align}
\pderiv{q_2}{t} &= \pderiv{\gz_2}{t} - \frac{k_R^2}{2}\pD{t}(\gy_2-\gy_1) \\
&= -J(\gy_2, \gz_2+f) + \frac{2f}{h}\gw + d_2 + e_2 - \frac{k_R^2}{2}\pD{t}(\gy_2-\gy_1) \\
&= -J(\gy_2, \gz_2+f) - \frac{k_R^2}{2}\left[\pD{t} (\gy_1-\gy_2) - J(\gy_1, \gy_2) - \frac{hQ_0}{2} \right] + d_2 + e_2 - \frac{k_R^2}{2}\pD{t}(\gy_2-\gy_1)\\
&= -J(\gy_2, \gz_2+f) + \frac{k_R^2}{2} J(\gy_1, \gy_2) + \frac{hk_R^2}{4}Q_0 + d_2 + e_2\\
&= -J(\gy_2, q_2) + Q + d_2 + e_2.
\end{align}
The governing equations Eq.~\eqref{eq:PotVortOne} and Eq.~\eqref{eq:PotVortTwo} for the potential vorticity follow.

\section{Streamfunction-forcing spectrum under random Gaussian forcing}
\label{app:StreamfunctionForcingSpectrum}

Let us consider the case of a generalized multi-layer model forced at each layer $\ga$ with random Gaussian forcing $f_\ga$ such that 
\begin{equation}
\avg{f_\ga (\bfx_1, t_1) f_\gb (\bfx_2, t_2)} = 2Q_{\ga\gb} (\bfx_1, \bfx_2) \gd (t_1-t_2).
\end{equation}
From the Novikov-Furutsu theorem \citep{article:Furutsu:1963,article:Novikov:1965} we know that, given a functional $R[f]$, the correlation between  $f_\ga$  and $R[f]$ reads
\begin{equation}
\avg{f_\ga (\bfx_1, t_1) R[f]} = \int_{\bbR^2} \df{\bfx_2}\int_\bbR \df{t_2} \; \avg{f_\ga (\bfx_1, t_1) f_\gb (\bfx_2, t_2)} \avg{\vderiv{R[f]}{f_\gb (\bfx_2, t_2)}}.
\end{equation}
It should be noted that implied is a space-time approach in which it is the entire forcing history $f$ that is being mapped to a number by the functional $R[f]$. The ensemble average is understood to average over all possible forcing histories. The idea is to treat the streamfunction $\gy_\ga$ of layer $\ga$ at a given point in space-time as a functional of the entire forcing history, and then use the Novikov-Furutsu theorem to evaluate the forcing-streamfunction spectrum  $\phi_{\ga\gb} (k)$. This idea follows a similar argument by \citet{book:McComb:1990}  for the three-dimensional Navier-Stokes equation. The argument proceeds as follows:

Recall first the definition of the filtering kernel: 
\begin{equation}
a^{<k}(\bfx, t) = \int_{\bbR^2} P(k|\bfx-\bfy) a(\bfy, t)\; \df{\bfy}.
\end{equation}
By definition, the streamfunction-forcing spectrum $\phi_{\ga\gb} (k)$ is given by
\begin{align}
\phi_{\ab}(k) &= \innerf{f_\ga}{\gy_\gb}{k} = \innerf{f_\ga}{\sum_\gc \ccL_{\bc}^{-1} q_\gc}{k} 
= \sum_\gc  L_{\bc}^{-1}(k) \innerf{f_\ga}{q_\gc}{k}\\
&= \sum_\gc  L_{\bc}^{-1}(k) \dD{k} \int_{\bbR^2}\df{\bfx}\; \avg{f_\ga^{<k} (\bfx,t) q_\gc^{<k} (\bfx,t)} \\
&= \sum_\gc  L_{\bc}^{-1}(k) \dD{k} \iiint_{(\bbR^2)^3}\df{\bfx}\df{\bfy}\df{\bfz}\;  P(k|\bfx-\bfy) P(k|\bfx-\bfz) \avg{f_\ga (\bfy,t) q_\gc (\bfz,t)}.
\label{eq:StreamfuncInProgress}
\end{align}
Using the Novikov-Furutsu theorem, we calculate the forcing-streamfunction correlation $\avg{f_\ga (\bfy,t) q_\gc (\bfz,t)}$, and find that it reads: 
\begin{align}
\avg{f_\ga (\bfy,t) q_\gc (\bfz,t)} &= \int_{\bbR^2}\df{\bfw}\int_\bbR\df{t_0} \; \avg{f_\ga (\bfy, t) f_{\gd} (\bfw,t_0)}\avg{\vderiv{q_\gc (\bfz, t)}{f_\gd (\bfw, t_0)}}\\
&= \int_{\bbR^2}\df{\bfw}\int_\bbR\df{t_0} \; 2Q_{\ad} (\bfy,\bfw)\gd (t-t_0) \avg{\vderiv{q_\gc (\bfz, t)}{f_\gd (\bfw, t_0)}}\\
&= 2\int_{\bbR^2}\df{\bfw} \; Q_{\ad} (\bfy,\bfw) \avg{\vderiv{q_\gc (\bfz, t)}{f_\gd (\bfw, t)}}.
\label{eq:StreamfuncCorrInProgress}
\end{align}
To evaluate the variational derivative of potential vorticity $q_\gc (\bfz, t)$ with respect to layer forcing $f_\gd (\bfw, t_0)$ we first note that, by causality, the potential vorticity $q_{\gc}(\bfz, t)$ at time $t$ is related with the initial potential vorticity $q_\gc (\bfz, 0)$ at time $t_0 = 0$ by an equation of the form
\begin{equation}
q_{\gc}(\bfz, t) = q_\gc (\bfz, 0)+\int_0^t \df{t_0}\; \cN_\gc [q(t_0)](\bfz) + \int_0^t \df{t_0}\; f_\gc (\bfz, t_0).
\end{equation}
Here, $\cN_\gc [q(t_0)](\bfz)$  represents the combined effect of the nonlinear and dissipation terms. The third integral represents the causal contribution of the forcing term. Let us assume now that $0< \gt < t$,  and differentiate the above equation variationally with respect to $f_\gd (\bfw, \gt)$. We immediately find that 
\begin{align}
\vderiv{q_{\gc}(\bfz, t)}{f_\gd (\bfw, \gt)} &= \vD{f_\gd (\bfw, \gt)} \left[ \int_0^t \df{t_0}\; \cN_\gc [q(t_0)](\bfz) + \int_0^t \df{t_0}\; f_\gc (\bfz, t_0) \right] \\
&= \cA_{\cd} (\bfz, t; \bfw, \gt) + \cB_{\cd} (\bfz, t; \bfw, \gt),
\end{align}
with $\cA_{\cd} (\bfz, t; \bfw, \gt)$  and $\cB_{\cd} (\bfz, t; \bfw, \gt)$  given by 
\begin{align}
\cA_{\cd} (\bfz, t; \bfw, \gt) &=  \vD{f_\gd (\bfw, \gt)} \int_0^t \df{t_0}\; \cN_\gc [q(t_0)](\bfz) = \int_{\gt}^t \df{t_0}\; \vderiv{\cN_\gc [q(t_0)](\bfz)}{f_\gd (\bfw, \gt)} \label{eq:TheAThing} \\
\cB_{\cd} (\bfz, t; \bfw, \gt) &= \vD{f_\gd (\bfw, \gt)}\int_0^t \df{t_0}\; f_\gc (\bfz, t_0)\\
&= \vD{f_\gd (\bfw, \gt)} \int_0^t \df{t_0}\int_{\bbR^2}\df{\bfz_0} \; \gd (\bfz-\bfz_0)f_\gc (\bfz_0, t_0) \\
&= \vD{f_\gd (\bfw, \gt)} \int_\bbR \df{t_0}\int_{\bbR^2}\df{\bfz_0} \; H(t-t_0) \gd (\bfz-\bfz_0)f_\gc (\bfz_0, t_0) \\
&= \gd_{\cd} H(\gt-t) \gd (\bfz-\bfw).
\end{align}
Here, $H(t)$ is the previously defined Heaviside function. For Eq.~\eqref{eq:TheAThing}, we rely on the principle of causality to restrict the integral from $\gt$  to $t$. 
It is easy to see that for $t=\gt$, $\cA_{\cd} (\bfz, t; \bfw, \gt)$ and $\cB_{\cd} (\bfz, t; \bfw, \gt)$ simplify to:
\begin{align}
\cA_{\cd} (\bfz, t; \bfw, t) &= \int_{t}^t \df{t_0}\; \vderiv{\cN_\gc [q(t_0)](\bfz)}{f_\gd (\bfw, \gt)} = 0,\\
\cB_{\cd} (\bfz, t; \bfw, t) &= \frac{1}{2} \gd_{\cd} \gd (\bfz-\bfw),
\end{align}
and therefore the variational derivative of $q_{\gc}(\bfz, t)$ with respect to $f_\gd (\bfw, t)$ is given by 
\begin{equation}
\vderiv{q_{\gc}(\bfz, t)}{f_\gd (\bfw, t)} =  \frac{1}{2} \gd_{\cd} \gd (\bfz-\bfw).
\end{equation}
Substituting this result to Eq.~\eqref{eq:StreamfuncCorrInProgress}, we show that the forcing-streamfunction correlation is given by 
\begin{align}
\avg{f_\ga (\bfy,t) q_\gc (\bfz,t)} &= 2\int_{\bbR^2}\df{\bfw} \; Q_{\ad} (\bfy,\bfw) \avg{\vderiv{q_\gc (\bfz, t)}{f_\gd (\bfw, t)}}\\
&= 2\int_{\bbR^2}\df{\bfw} \; Q_{\ad} (\bfy,\bfw) \frac{1}{2} \gd_{\cd} \gd (\bfz-\bfw)\\
&= \int_{\bbR^2}\df{\bfw} \; Q_{\ac} (\bfy,\bfw) \gd (\bfz-\bfw) = Q_{\ac} (\bfy,\bfz).
\end{align}
Consequently, the forcing-streamfunction spectrum  $\phi_{\ga\gb} (k)$ reads 
\begin{align}
\phi_{\ab}(k) &= \sum_\gc  L_{\bc}^{-1}(k) \dD{k} \iiint_{(\bbR^2)^3}\df{\bfx}\df{\bfy}\df{\bfz}\;  P(k|\bfx-\bfy) P(k|\bfx-\bfz) \avg{f_\ga (\bfy,t) q_\gc (\bfz,t)}\\
&= \sum_\gc  L_{\bc}^{-1}(k) \dD{k} \iiint_{(\bbR^2)^3}\df{\bfx}\df{\bfy}\df{\bfz}\;  P(k|\bfx-\bfy) P(k|\bfx-\bfz) Q_{\ac} (\bfy,\bfz)\\
&= \sum_\gc \cQ_{\ac}(k) L_{\bc}^{-1}(k).
\end{align}
The integral above defines the forcing correlation spectrum $\cQ_{\ac}(k)$, given by:
\begin{equation}
\cQ_{\ac}(k) = \dD{k} \iiint_{(\bbR^2)^3}\df{\bfx}\df{\bfy}\df{\bfz}\;  P(k|\bfx-\bfy) P(k|\bfx-\bfz) Q_{\ac} (\bfy,\bfz).
\end{equation}
Our final result for the forcing-streamfunction spectrum is:
\begin{equation}
\phi_{\ab}(k) = \sum_\gc \cQ_{\ac}(k) L_{\bc}^{-1}(k).
\end{equation}

\section{The random Gaussian forcing model}
\label{app:RandomGaussianForcing}

Let us consider the case of the two-layer quasigeostrophic model forced with $f_1=\gf$  at the top layer and $f_2=-\mu \gf$  at the bottom layer. Here, $\mu$ is a suppression constant with $0 < \mu  < 1$ and $\gf$ is a random Gaussian field that is delta-correlated in time such that:
\begin{equation}
\avg{\gf (\bfx_1, t_1) \gf (\bfx_2, t_2)} = 2Q (\bfx_1, \bfx_2) \gd (t_1-t_2).
\end{equation}
From $Q (\bfx_1, \bfx_2)$  we define the corresponding correlation spectrum $\cQ(k)$ as:
\begin{equation}
\cQ(k) = \dD{k} \iiint_{(\bbR^2)^3}\df{\bfx}\df{\bfy}\df{\bfz}\;  P(k|\bfx-\bfy) P(k|\bfx-\bfz) Q (\bfy,\bfz).
\end{equation}
It follows that, for $\ga,\gb \in \{1,2\}$, $f_\ga$ and $f_\gb$  are correlated according to 
\begin{equation}
\avg{f_\ga (\bfx_1, t_1) f_\gb (\bfx_2, t_2)} = 2Q_{\ab} (\bfx_1, \bfx_2) \gd (t_1-t_2),
\end{equation}
with the components of $Q_{\ab}$  given by 
\begin{align}
Q_{11}(\bfx_1, \bfx_2) &= Q(\bfx_1, \bfx_2),\\
Q_{12}(\bfx_1, \bfx_2) &= Q_{21}(\bfx_1, \bfx_2) = -\mu Q(\bfx_1, \bfx_2),\\
Q_{22}(\bfx_1, \bfx_2) &= \mu^2 Q(\bfx_1, \bfx_2).
\end{align}
The spectrum $\cQ_{\ab}(k)$  of $Q_{\ab} (\bfx_1, \bfx_2)$  is defined as
\begin{equation}
\cQ_{\ab}(k) = \dD{k} \iiint_{(\bbR^2)^3}\df{\bfx}\df{\bfy}\df{\bfz}\;  P(k|\bfx-\bfy) P(k|\bfx-\bfz) Q_{\ab} (\bfy,\bfz),
\end{equation}
consequently its components read:
\begin{align}
\cQ_{11}(k) &= \cQ(k), \\
\cQ_{12}(k) &= \cQ_{21}(k) = -\mu \cQ(k), \\
\cQ_{22} (k)&= \mu^2 \cQ(k).
\end{align}
In Appendix~\ref{app:StreamfunctionForcingSpectrum}, we have shown that under general delta-correlated in time random Gaussian forcing, the general form of the forcing-streamfunction spectrum $\phi_{\ga\gb} (k)$ reads:
\begin{equation}
\phi_{\ab}(k) = \sum_\gc \cQ_{\ac}(k) L_{\bc}^{-1}(k).
\end{equation}
We would now like to reduce this result to the case of the two-layer quasi-geostrophic model. Starting from Eq.~\eqref{eq:TheMatrixL}, a simple calculation shows that the inverse matrix $L_{\ab}^{-1}(k)$ is given by:
\begin{equation}
L_{\ab}^{-1}(k) = \frac{-1}{2k^2 (k^2+k_R^2)} \mattwo{2k^2+k_R^2}{k_R^2}{k_R^2}{2k^2+k_R^2}.
\end{equation}
We note that the inverse matrix $L_{\ab}^{-1}(k)$ is defined for all wavenumbers $k>0$. Combining the above two equations  we find that the components of the streamfunction-forcing spectrum $\phi_{\ga\gb} (k)$ are:
\begin{align}
\phi_{11}(k) &= \cQ_{11}(k)L^{-1}_{11}(k) + \cQ_{12}(k)L^{-1}_{12} (k)
= \cQ(k) [L_{11}^{-1}(k)-\mu L_{12}^{-1}(k)] \\
&= \frac{-\cQ(k)[2k^2+k_R^2-\mu k_R^2]}{2k^2 (k^2+k_R^2)}, \\ 
\phi_{12}(k) &=   \cQ_{11}(k)L_{21}^{-1}(k) +  \cQ_{12}(k)L_{22}^{-1} (k)
= \cQ(k) [L_{21}^{-1}(k)-\mu L_{22}^{-1}(k)] \\
&= \frac{-\cQ(k)[k_R^2-\mu (2k^2+k_R^2)]}{2k^2 (k^2+k_R^2)}, \\
\phi_{21}(k) &=  \cQ_{21}(k)L_{11}^{-1}(k) +  \cQ_{22}(k)L_{12}^{-1}(k) 
= \cQ(k) [-\mu L_{11}^{-1}(k)+\mu^2 L_{12}^{-1}(k) ]\\
&= \frac{-\cQ(k)[-\mu (2 k^2+k_R^2) + \mu^2 k_R^2]}{2k^2 (k^2+k_R^2)}, \\
\phi_{22}(k) &=  \cQ_{21}(k)L_{21}^{-1}(k) +  \cQ_{22}(k)L_{22}^{-1}(k) 
= \cQ(k) [-\mu L_{21}^{-1}(k)+\mu^2  L_{22}^{-1}(k)] \\
&= \frac{-\cQ(k)[-\mu k_R^2 + \mu^2 (2k^2+k_R^2)]}{2k^2 (k^2+k_R^2)}. 
\end{align}
We may therefore write the streamfunction-forcing spectra as:
\begin{equation}
\gf_{\ab} (k) = \frac{-\cQ (k) \gy_{\ab} (k)}{2 k^2 (k^2+k_R^2)},
\end{equation}
with $\gy_{\ab}$ given by
\begin{align}
\gy_{11} (k)  &= (2k^2+k_R^2)-\mu k_R^2, \\
\gy_{12} (k)  &= k_R^2-\mu (2k^2+k_R^2), \\
\gy_{21} (k)  &= -\mu (2 k^2+k_R^2) + \mu^2 k_R^2, \\
\gy_{22} (k)  &= -\mu k_R^2 + \mu^2 (2k^2+k_R^2).
\end{align}
From the streamfunction-forcing spectra $\phi_{\ga\gb} (k)$ we calculate both the potential enstrophy forcing spectrum $F_G(k)$ and the energy forcing spectrum $F_E(k)$  using Eq.~\eqref{eq:FE} and  Eq.~\eqref{eq:FG}. An easy calculation gives:
\begin{align}
\psi_{11}(k) + \psi_{22}(k) &= (2k^2+k_R^2)-\mu k_R^2-\mu k_R^2 + \mu^2 (2k^2+k_R^2)\\
&= 2(1+\mu^2)k^2 + k_R^2 (1-\mu)^2,
\end{align}
and therefore,  for $F_E(k)$ we find that
\begin{align}
F_E (k) &= -2[\phi_{11}(k) + \phi_{22}(k)] = \frac{+2\cQ(k)[\psi_{11}(k) + \psi_{22}(k)]}{2k^2 (k^2+k_R^2)}\\
&= \frac{2Q(k)[2(1+\mu^2) k^2 + (1-\mu)^2 k_R^2]}{2k^2 (k^2+k_R^2)}.\label{eq:theFE}
\end{align}
For  the potential enstrophy forcing spectrum $F_G(k)$, we use a slightly more subtle argument, and we have:
\begin{align}
F_G (k) &= 2\sum_{\ga\gb} L_{\ab}(k) \phi_{\ab}(k)
=  2\sum_{\ga\gb} L_{\ab}(k) \left[ \sum_\gc L_{\bc}^{-1} (k) \cQ_{\ac}(k) \right]\\
&=  2\sum_{\ac} \left[ \sum_\gb L_{\ab}(k) L_{\bc}^{-1} (k) \right]\cQ_{\ac}(k)
= 2\sum_{\ac} \gd_{\ac}\cQ_{\ac}(k)   \\
&= 2\sum_{\ga} \cQ_{\ga\ga}(k) 
= 2[\cQ_{11}(k) + \cQ_{22}(k)] \\ 
&= 2(1+\mu^2) \cQ(k). \label{eq:theFG}
\end{align}
It is worth noting that the potential enstrophy forcing spectrum $F_G(k)$ is independent of the matrix  $L_{\ab}(k)$  as long as  $L_{\ab}(k)$  is non-singular. The energy forcing spectrum $F_E(k)$, on the other hand, is dependent on the inverse matrix $L_{\ab}^{-1}(k)$. Eq.~\eqref{eq:theFE} and Eq.~\eqref{eq:theFG} are the main results of this appendix.

\bibliography{references}
\bibliographystyle{jfm}

\end{document}